\documentclass[structabstract]{aa}
\usepackage{graphicx}
\usepackage{natbib} \bibpunct{(}{)}{;}{a}{}{,}
\usepackage{color}
\usepackage{rotating}
\usepackage{txfonts}

\def\remark{}
\def\changed{}
 
 \newcommand{\msunpyr}{\,M_\odot\,\mbox{yr}^{-1}}
\newcommand{\dint}{\,\mbox{d}} 

\newcommand{\kms}{\ifmmode{\,\mbox{km}\,\mbox{s}^{-1}}\else{km/s}\fi}
\newcommand{\msun}{\ifmmode M_{\odot} \else M$_{\odot}$\fi}
\newcommand{\rsun}{\ifmmode R_{\odot} \else R$_{\odot}$\fi}
\newcommand{\lsun}{\ifmmode L_{\odot} \else L$_{\odot}$\fi}
\newcommand{\zsun}{\ifmmode Z_{\odot} \else $Z_{\odot}$\fi}
\newcommand{\velo}{\ifmmode\varv\else$\varv$\fi}
\newcommand{\vinf}{\ifmmode\velo_\infty\else$\velo_\infty$\fi}
\begin{document} 
 
\title{On the optically-thick winds of Wolf-Rayet stars}

\author{G.\ Gr\"{a}fener\inst{\ref{inst1},\ref{inst2}} 
\and S.P.\ Owocki\inst{\ref{inst3}}
\and L.\ Grassitelli\inst{\ref{inst1}} 
\and N.\ Langer\inst{\ref{inst1}} 
}
 
\institute{Argelander-Institut f{\"u}r Astronomie der Universit{\"a}t Bonn, Auf dem H{\"u}gel 71, 53121 Bonn, Germany\label{inst1}
\and Armagh Observatory, College Hill, Armagh BT61\,9DG, United Kingdom\label{inst2}
\and Bartol Research Institute, University of Delaware, Newark, DE 19716, USA\label{inst3}
}

\date{Received ; Accepted}

\abstract{The classical Wolf-Rayet (WR) phase is believed to mark the
  end stage of the evolution of massive stars with initial masses
  higher {\changed than $\sim\,25\,M_\odot$}. Stars in this phase
  expose their stripped cores with the products of H- or He-burning at
  their surface. They develop strong, optically-thick stellar winds,
  which are important for the mechanical and chemical feedback of
  massive stars, and which determine whether the most massive stars
  end their lives as neutron stars or black holes. Currently the winds
  of WR stars are not well understood and their inclusion in stellar
  evolution models relies on uncertain empirical mass-loss relations.}
{In this work we investigate theoretically the mass-loss properties of
  H-free WR stars of the nitrogen sequence (WN stars).}
{We connect stellar structure models for He stars with wind models for
  optically-thick winds and assess how both types of models can
  simultaneously fulfill their respective {\changed sonic-point
    conditions}.}
{Fixing the outer wind law and terminal wind velocity $\varv_\infty$,
  we obtain unique solutions for {\changed the mass-loss rates of
    optically-thick, radiatively-driven winds of WR stars in the phase
    of core He-burning}. The resulting mass-loss relations as a
  function of stellar parameters, agree well with previous empirical
  relations. Furthermore, we encounter {\changed stellar mass limits
    below which no continuous solutions exist.} While these mass
  limits agree with {\changed observations} of WR stars in the Galaxy,
  they are in conflict with {\changed observations in the LMC}.}
{While our results confirm in particular the slope of oft-used
  empirical mass-loss relations, they imply that only part of the
  observed WN population can be understood in the framework of the
  standard assumptions of a smooth transonic flow and compact stellar
  core. This means that alternative approaches, such as a clumped and
  inflated wind structure, or deviations from the diffusion limit at
  the sonic point may have to be invoked.  Qualitatively, the
  existence of mass limits for the formation of WR-type winds may be
  relevant for the non-detection of low-mass WR stars in binary
  systems, which are believed to be progenitors of Type Ib/c
  supernovae. The sonic-point conditions derived in this work may
  provide a possibility to include optically-thick winds in stellar
  evolution models in a more physically motivated form than in current
  models.}

\keywords{Stars: Wolf-Rayet -- Stars: early-type -- Stars: atmospheres
  -- Stars: mass-loss -- Stars: winds, outflows}
\maketitle

\section{Introduction}

The {\changed strong} winds of emission-line stars like Wolf-Rayet
(WR) stars or Luminous Blue Variables (LBVs) are of key importance for
the mechanical and chemical feedback of massive stars on the
interstellar medium. They decide how much mass and angular momentum
massive stars lose before their final collapse, and whether the most
massive stars end their lives as neutron stars or black holes.

{\changed The occurrence of emission lines in the spectra of WR stars
  and LBVs is related to the large optical thickness of their winds.
  First of all, this means that, at short wavelengths, ionising
  photons are efficiently absorbed by optically-thick continua within
  the extended wind. The subsequent recombination then leads to the
  observed emission-line cascades at longer wavelengths. Secondly,
  this means that photons can be absorbed and re-emitted several times
  before they escape the wind \citep{luc1:93}. This way the radiation
  field can transfer a multiple of its photon momentum to the wind
  material, leading to wind-efficiency numbers
  $\eta = \dot{M}\varv_\infty/(L/c) > 1$, where $\eta$ denotes the
  ratio between the wind momentum {\remark rate} (the product of
  mass-loss rate $\dot{M}$ and terminal wind velocity $\varv_\infty$)
  and the momentum of the radiation field (the stellar luminosity $L$
  divided by the speed of light $c$). As we will discuss later in this
  work, this also means that the sonic point of optically-thick winds
  is located at flux-mean optical depths $\tau_{\rm s}>1$
  \citep[cf.][]{vin1:12}.}

{\changed In the limit of large optical depth $\tau_{\rm s}\gg 1$, and
  for sufficiently small velocity gradients (cf.\ Eq.\ref{eq:CAK}),
  the physics of radiation-driven winds simplifies because the
  radiative transfer can be described in the diffusion limit, and the
  flux-mean opacity $\kappa_F$ equals the Rosseland-mean opacity
  {\remark $\kappa_{\rm R}(\rho,T)$} which is a function of density
  $\rho$ and temperature $T$ only. In this limit the equation of
  motion has a critical point at the sonic point that allows to infer
  some properties of winds in this regime. The critical conditions
  arising in this limit for the winds of WR stars have been
  investigated by \citet{nug1:02}.}

For WR stars \citet{lam1:02} distinguished between hot and cool winds
depending on the temperature at their sonic point. They identified a
hot regime with sonic-point temperatures $\gtrsim$\,160\,kK slightly
above the temperature of the Fe-opacity peak, and a cool regime with
temperatures of 40--70\,kK related to a cooler opacity bump. Detailed
hydrodynamic atmosphere/wind models for hot WR stars of the carbon
sequence \citep[WC stars,][]{gra1:05} and cool H-rich WR stars of the
nitrogen sequence \citep[WNh stars,][]{gra1:08} confirmed these
temperature ranges although there may be an ambiguity between
different opacity sources due to He, C, and Fe in the cool temperature
range\footnote{\changed while \citet{nug1:00,ro1:16} identify He and C
  (the latter in carbon-rich mixtures) as the main opacity source,
  \citet{lam1:02,gra1:08} identify Fe as the main contributor. In the
  non-LTE models of \citeauthor{gra1:08} these differences are most
  likely caused by the moderate wind optical depths of WNh stars, and
  the consequent break-down of the diffusion approximation in this
  regime.}.

These studies further showed that it is possible to drive the outer
part of optically-thick, WR-type winds by radiation pressure alone,
and that optically-thick winds are formed when stars are approaching
the Eddington limit. The resulting dependence {\changed of the
  mass-loss rate} on the classical Eddington factor
\footnote{$\Gamma_{\rm e} = \kappa_{\rm e} L / (4\pi c G M)$ denotes
  the Eddington factor defined with respect to the free-electron
  opacity $\kappa_{\rm e}$.} $\Gamma_{\rm e}$ has been confirmed by
means of Monte-Carlo models of \citet{vin1:11} who identified a kink
in the $\Gamma_{\rm e}$-dependence between the optically-thick and
optically-thin regime.

Indeed, WR-type winds are found for a variety of objects with high
Eddington factors mainly due to their high $L/M$ ratios. Evolved
{\changed massive} stars, such as H-free WN and WC stars, {\changed
  reach $L/M$ ratios of the order of $10^4\,L_\odot/M_\odot$} due to
the enhanced mean molecular weight in their He-burning core. [WC]-type
central stars of planetary nebulae {\changed reach} high $L/M$ ratios
as a consequence of shell burning on top of their {\changed
  electron-}degenerate core \citep[cf.][]{gra5:08}. WNh stars, on the
other hand, are main-sequence stars with masses
$\gtrsim$\,$100\,M_\odot$. They {\changed reach high $L/M$ ratios}
because $L$ increases much more steeply than $M$ for increasing
stellar masses.

A related effect occurring near the Eddington limit is the envelope
inflation effect \citep{ish1:99,pet1:06,gra1:12,san1:15}. This effect
leads to a radial extension of the outer stellar envelope {\changed
  with a density inversion \citep{jos1:73}}, and may be responsible
for the large radii that are deduced in spectroscopic analyses of WR
stars \citep[e.g.][]{ham1:06,san1:12,hai1:14}. The discrepancy between
the empirical radii of H-free WR stars and the ones expected for stars
on the He-main sequence can amount up to a factor 10, and is known as
the WR radius problem. \citet{gra1:12} found that an enhanced mean
opacity due to an inhomogeneous (clumped) structure within the
inflated envelope can {\changed possibly} account for the large
observed radii. The presence of inhomogeneities in similar,
radiation-dominated envelopes is supported by 3-dimensional models of
\citet{jia1:15}, {\changed however, these authors argue that the
  effects of porosity in an inhomogeneous medium can reduce the mean
  opacity \citep[cf.][]{sha1:98,owo1:04,osk1:07,owo1:08}}.

Based on a semi-empirical analysis \citet{gra2:13} investigated the
sonic-point conditions that follow from the observed properties of the
outer winds of WC stars. They found sonic-point temperatures and
densities that would imply enhanced opacities for the majority of
stars, which may be the consequence of a clumped envelope
structure. Their results suggested that a small number of stars may
have smooth envelopes with a compact structure and a hot sonic-point
temperature, while the majority has clumped envelopes with an inflated
structure and cool sonic-point temperatures.

It is presently not entirely clear how such a clumped and inflated
structure can be realised in nature. In many cases an inflated
envelope structure would suggest densities that are so low that the
velocities within the envelope, due to the equation of continuity,
would reach values comparable to those in a stellar wind {\changed
  \citep[cf.][]{pet1:06,ro1:16}. Except for very low mass-loss rates
  the models in both works have small sonic-point radii suggesting a
  compact rather than an inflated envelope structure.

  To reconcile these results with the radii derived from spectroscopic
  analyses it may be necessary to invoke more exotic scenarios for the
  layers above the sonic point. In fact \citet{pet1:06,ro1:16} both
  find that the wind acceleration ceases beyond the Fe-opacity
  bump. Also \citet{gra1:05} reported problems with the wind
  acceleration in the same region.}  It may thus be that inflated
envelopes have more in common with a failed wind, or that there are
multiple components of material which are partly falling back on the
star.

In the present work we intend to test how the assumption of a smooth,
hot/compact stellar wind complies with the observed properties of
H-free WN stars. This type of star has the advantage that it is well
described by pure He models that contain trace elements beyond He in
solar-like abundances. Even if their cores are enriched by the
products of He-burning the mean molecular weight of these stars stays
almost the same. This means that their masses and in particular their
$L/M$ ratios are well determined by a given (observed) luminosity
{\changed \citep{lan1:89}}.

In the following we construct models for the winds of hot/compact He
stars, assuming that they are optically thick and radiatively driven.
To this purpose we combine the critical conditions arising at the
sonic point, with additional conditions imposed by the outer wind,
assuming a fixed outer wind structure.  In Sect.\,\ref{sec:winds} we
explain our method and discuss how our results depend on the adopted
input parameters. In Sect.\,\ref{sec:galwn} we apply our models to a
comprehensive sample of H-free WN stars in the Galaxy and the Large
Magellanic Cloud (LMC). In Sect.\,\ref{sec:discussion} we discuss the
results.

\section{Optically-thick winds}
\label{sec:winds}

{\changed As indicated in the previous section, radiation-driven winds
  can be divided into} optically-thin and optically-thick winds,
depending on the location of their critical point. For the thin winds
of OB stars \citet[][CAK]{cas1:75} formulated a closed theoretical
description that allows to predict wind parameters, such as the
mass-loss rate $\dot{M}$ and terminal wind velocity $\varv_\infty$,
for a given set of stellar parameters, such as the stellar mass $M$,
luminosity $L$, and radius $R$. Within the CAK theory the wind
parameters are determined using so-called force multipliers, i.e.,
parameters that characterise the strength and the distribution of the
spectral lines that drive the wind.  Line driving is particularly
efficient in optically-thin winds because the Doppler shifts that
arise in the accelerated wind expose spectral lines to an unattenuated
radiation field, and reduce the effects of line self-shadowing.

In optically-thin winds these Doppler shifts introduce a dependency of
the radiative line acceleration on the velocity gradient
$\partial \varv / \partial r$. This term alters the equation of motion
in a way that a critical point arises at the velocity of a
radiative-acoustic wave mode that is faster than the sonic speed
\citep[the so-called Abbott speed, cf.][]{abb1:80}. Moreover, the
dependence on $\partial \varv / \partial r$ enables a simultaneous
determination of the mass-loss rate {\em and} wind velocity from the
conditions arising at the critical point.

When winds become optically thick, the effect of Doppler shifts
becomes less efficient, because the photon mean-free path decreases.
Whether Doppler shifts are important or not depends on the {\changed
  CAK optical-depth parameter \citep{cas1:75}
\begin{equation}
\label{eq:CAK}
  t_{\rm CAK} = \frac{\kappa_{\rm e} \rho \varv_{\rm Dop}}{\partial \varv/ \partial r}
\end{equation}
which denotes the ratio between the Sobolev length
($\varv_{\rm Dop}/(\partial \varv/ \partial r)$) and the photon
mean-free path ($1/\kappa_{\rm e} \rho$). Here $\kappa_{\rm e}$
denotes the reference continuum opacity due to free electrons, and
$\varv_{\rm Dop}$ the line-broadening velocity due to {\remark thermal
  and turbulent gas motions}.  For WR stars \citet{nug1:02} estimated
values up to $t_{\rm CAK} \approx 50$--150 at the sonic point, which
would suggest that Doppler shifts may indeed not affect the
sonic-point conditions (but see our discussion in
Sect.\,\ref{sec:disc_winds}). In the remainder of this work we assume
that the wind optical depth is large enough so that Doppler shifts can
be neglected.}

In the optically-thick limit the radiation field is better described
by photon diffusion. In this limit photons tend to avoid strong lines
and choose the path of lowest resistance through gaps between lines in
frequency space. For this reason it is much more difficult to launch
and accelerate optically-thick winds, such as the winds of WR stars or
LBVs.

In the diffusion limit, and assuming atomic level populations in local
thermodynamic equilibrium (LTE), the radiative acceleration
$g_{\rm rad}$ can be expressed much more easily, by the product of the
Rosseland-mean opacity $\kappa_{\rm R}$ and the radiative flux $F$
divided by the speed of light $c$.
\begin{equation}
  \label{eq:grad}
  g_{\rm rad} = \kappa_F \times \frac{F}{c} = \kappa_{\rm R}(\rho, T) \times \frac{F}{c}.
\end{equation}
In this equation the first equality denotes the general case with the
flux-mean opacity $\kappa_F$, and the second equality the diffusion
limit where $\kappa_F$ can be expressed as a function of density and
temperature via the Rosseland-mean opacity $\kappa_{\rm R}$. In the
following we discuss the sonic-point conditions arising in this limit
in Sect.\,\ref{sec:sonic} and the mass-loss relations that follow from
a connection between stellar wind and envelope at the sonic point in
Sect.\,\ref{sec:connect}.

\subsection{Sonic-point conditions}
\label{sec:sonic}

The sonic-point conditions for radiatively-driven winds in the
diffusion limit have been discussed in detail by
\citet{nug1:02,ro1:16}.  Most importantly, the equation of motion has
a critical point at the sonic radius $R_{\rm s}$ where the velocity
$\varv(r)$ equals the isothermal sound speed $a$.

According to \citet{nug1:02} this critical condition can be expressed
in very good approximation as
\begin{equation}
\label{eq:crit}
  \Gamma(R_{\rm s})\equiv\frac{\kappa_{\rm R}(\rho_{\rm s}, T_{\rm s})L}{4\pi cGM}
  \simeq 1,
\end{equation}
where $\rho_{\rm s}$ and $T_{\rm s}$ denote the density and
temperature at the sonic point. This means that the wind solution has
to cross the Eddington limit very close to the sonic radius
$R_{\rm s}$.  Furthermore, $\kappa_{R}$ needs to increase with radius,
to ensure that the wind can be accelerated through the sonic point.

For their wind models \citet{nug1:02} adopted $\Gamma(R_{\rm s}) = 1$,
or equivalently,
$\kappa_{\rm R}(\rho_{\rm s}, T_{\rm s}) = \kappa_{\rm Edd}$ with the
Eddington opacity $\kappa_{\rm Edd} = 4\pi c G M /L$. This means that,
in principle, the sonic-point conditions can be extracted directly
from opacity tables where $\kappa_{\rm R}$ is given as a function of
$\rho$ and $T$.

Based on dynamical wind models \citet{ro1:16} have shown that the
subsonic density and temperature structure of their WR models almost
precisely matches the structure of {\changed hydrostatic} models from
\citet{gra1:12}. For the parameter range that we are interested in,
also these models approach $\Gamma=1$ near the Fe-opacity peak, i.e.,
{\changed dynamical terms do not significantly affect the subsonic
  structure. Therefore} we take advantage of this situation and use
the density and temperature from {\changed hydrostatic models} to
determine the sonic-point conditions, instead of employing {\changed
  more complex hydrodynamical} models as the ones by {\changed
  \citet{pet1:06,ro1:16}}.

For our analysis we use the gas pressure $P_{\rm gas}$ and radiation
pressure $P_{\rm rad}$ as independent variables (instead of density
and temperature) to describe the conditions at the sonic point. As, by
definition, $\varv=a$ at the sonic point and thus
$\dot{M} = 4\pi R_{\rm s}^2 \rho_{\rm s} a_{\rm s}$ we can express
$P_{\rm gas}$ and $P_{\rm rad}$ at the sonic point, for a given
mass-loss rate $\dot{M}$ and sonic radius $R_{\rm s}$, as
\begin{equation}
  \label{eq:pgas}
  P_{\rm gas}=\rho_{\rm s} a_{\rm s}^2 = \frac{\dot{M}a_{\rm s}}{4\pi R_{\rm s}^2}
  = \frac{\dot{M}}{4\pi R_{\rm s}^2} 
  \times \left(\frac{{\cal R} T_{\rm s}}{\mu}\right)^{\frac{1}{2}},
\end{equation}
where we used the definition $a^2={{\cal R} T}/{\mu}$ with the gas
constant ${\cal R}$ and mean molecular weight $\mu$, and
\begin{equation}
  P_{\rm rad}= \frac{4\sigma}{3c} T_{\rm s}^4,
\end{equation}
with the Stefan-Boltzmann constant $\sigma$ and speed of light $c$.

This leads to an expression for $P_{\rm gas}$ as a function of
$\dot{M}$, $R_{\rm s}$, and $P_{\rm rad}$ at the sonic point
\begin{equation}
P_{\rm gas}^8
= \left(\frac{\dot{M}}{4\pi R_{\rm s}^2}\right)^8
  \left( \frac{{\cal R}}{\mu}\right)^4 \frac{3c}{4\sigma} P_{\rm rad}
\end{equation}
or
\begin{equation}
\label{eq:mdot}
\log(P_{\rm gas})
= \log\left[\frac{\dot{M}/(\msunpyr)}{(R_{\rm s}/R_\odot)^2}\right] +
  8.737 + \log(P_{\rm rad})/8,
\end{equation}
with $P_{\rm rad}$ and $P_{\rm gas}$ in cgs units and $\mu=4/3$ for an
ionised He plasma.

\begin{figure}[tbp!]
  \parbox[b]{0.49\textwidth}{\center{\includegraphics[scale=0.35]{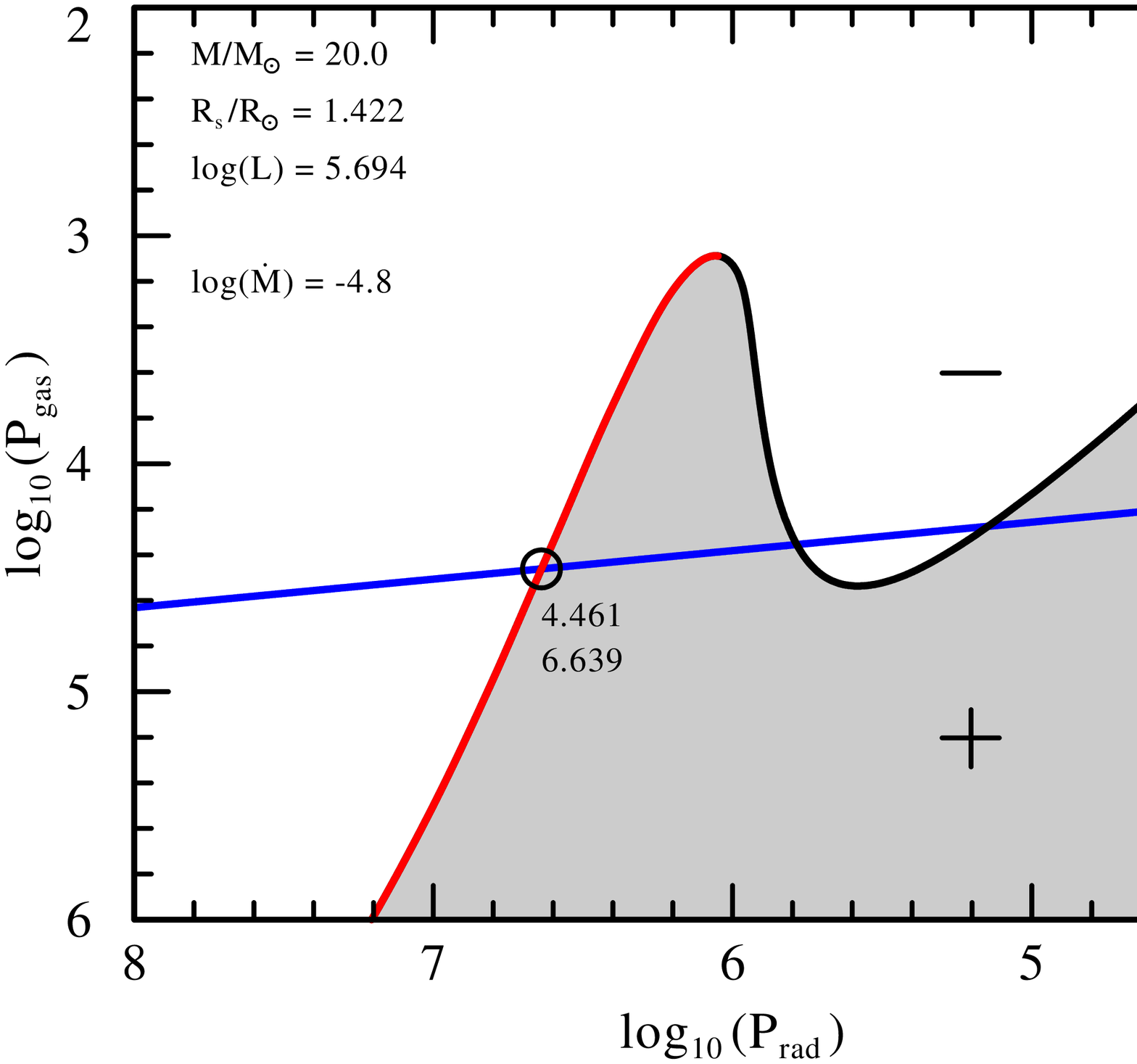}}}
  \caption{{\remark Critical conditions in $P_{\rm
        rad}$--$P_{\rm
        gas}$
      plane. The hydroststic envelope solution for a $20\,M_\odot$ He
      star from \citet{gra1:12} is indicated by the black/red curve
      (the axes in this diagram are inverted so that the stellar
      interior is located in the lower left and the stellar surface in
      the upper right).  On this curve
      $\kappa(P_{\rm rad}, P_{\rm gas})\approx \kappa_{\rm Edd}$.}
    Areas in the $P_{\rm rad}$--$P_{\rm gas}$ plane with higher/lower
    opacities are indicated in grey/white {\changed and by $+/-$
      signs}. The blue line indicates possible locations of the sonic
    point {\remark for an adopted mass-loss rate of
      $10^{-4.8}\msunpyr$} following Eq.\,\ref{eq:mdot}. The circle
    indicates the innermost intersection point with an
    outward-increasing opacity, giving a precise estimate for the
    location of the sonic point (see text).}
  \label{fig:CRIT}
\end{figure}

In Fig.\,\ref{fig:CRIT} it is illustrated how, for a $20\,M_\odot$ He
star with solar metallicity ($Z=0.02$)\footnote{\changed the models
  are identical with the stellar structure models for He stars from
  \citet{gra1:12}, which are based on OPAL Rosseland-mean opacity
  tables \citep{igl1:96} with solar-like trace element abundances
  \citep{gre1:93}.} {\changed and a given mass-loss rate}, the
intersection point between the static envelope solution and
Eq.\,\ref{eq:mdot} determines the values of $P_{\rm rad}$ and
$P_{\rm gas}$ that a dynamical model would {\changed have} at the
sonic point. The envelope model in Fig.\,\ref{fig:CRIT} follows an
S-shaped solution with high $P_{\rm rad}$ and $P_{\rm gas}$ in the
stellar interior and low $P_{\rm rad}$ and $P_{\rm gas}$ at the
stellar surface. As pointed out before, the stellar envelope is
dominated by radiation pressure, and follows very closely a contour
with $\Gamma=1$ in the $P_{\rm rad}$--$P_{\rm gas}$ plane as long as
the ratio $P_{\rm rad}/P_{\rm gas}$ is high. The S-shape of the
envelope solution is a consequence {\changed of} the topology of the
opacity in the $P_{\rm rad}$--$P_{\rm gas}$ plane, and in particular
of the presence of the Fe opacity peak near
$\log(P_{\rm rad})\approx 6$ in cgs units.

In Fig.\,\ref{fig:CRIT}, the envelope solution and Eq.\,\ref{eq:mdot}
have three intersection points which could in principle become the
sonic point of an optically-thick wind. However, only for the inner-
and outermost intersections the opacity increases towards the stellar
surface. These two points represent the hot and cool wind solutions
discussed by \citet{nug1:02}.

In the following we concentrate on the hot solution branch which is
indicated in red in Fig.\,\ref{fig:CRIT}. For these solutions the
stellar envelope is compact and the sonic point lies at hot
temperatures, below the Fe opacity peak with $P_{\rm rad}\gg P_{\rm
  gas}$.

\subsection{Connection of wind and envelope}
\label{sec:connect}

To constrain the mass-loss rates of WR stars we further take the
sonic-point conditions into account that are imposed by the outer
wind. Such a connection of wind and envelope has been discussed in a
qualitative manner by \citet{gra2:13}.

The underlying idea of their approach was to estimate the temperature
at the sonic point of a radiatively-driven wind {\changed based on} a
prescribed velocity structure $\varv(r)$ assuming that it is
radiatively driven, i.e., using the opacities that are needed to
accelerate the wind. {\changed Based on numerical computations they
  found a relation between $P_{\rm gas}$ and $P_{\rm rad}$ (or
  alternatively $\rho$ and $T$) at the sonic point which is imposed by
  the back-warming effect of the outer wind}, namely that
$P_{\rm rad}/P_{\rm gas} \approx const$. In Sect.\,\ref{sec:prpg} we
describe our approach to compute this {\changed new} 'wind condition'.

In Sect.\,\ref{sec:eta} we elaborate on an approximate approach to
estimate the wind condition analytically. While the applicability of
this result may be limited, the derivations in this section help to
understand why the wind condition results in relations with an almost
constant ratio of
$P_{\rm rad}/P_{\rm gas} \approx \varv_\infty/a_{\rm s}$. We further
combine the approximate wind condition with the critical condition
(Eq.\,\ref{eq:crit}) to estimate mass-loss rates for WR stars.

In Sect.\,\ref{sec:numerics} we evaluate the wind condition from
Sect.\,\ref{sec:prpg} numerically, and derive improved mass-loss
estimates. We further discuss how the numerical relations for
$P_{\rm rad}$ vs. $P_{\rm gas}$ relate to the results from
Sect.\,\ref{sec:eta}. Finally, we discuss in
Sect.\,\ref{sec:mdot_vinf} how the derived mass-loss rates depend on
the dominant input parameter of our models, the terminal wind velocity
$\varv_\infty$.

\subsubsection{$P_{\rm rad}$ and $P_{\rm gas}$ at the sonic point}
\label{sec:prpg}

{\changed For a given outer wind structure}
the value of $P_{\rm rad}$, or equivalently the temperature at the
sonic point, is a consequence of the back-warming effect of the
optically-thick wind. To compute $P_{\rm rad}$ we use the same
formalism as \citet[][cf. their Sects.\ 3.1 and 3.2 for more
details]{gra2:13} which is based on the original works by
\citet{luc1:71,luc1:76,luc1:93}.

$P_{\rm rad}$ is connected to the radiative acceleration $g_{\rm rad}$
(Eq.\,\ref{eq:grad}) via
\begin{equation}
\label{eq:PRGR}
\frac{\dint P_{\rm rad}}{\dint r} = - \rho g_{\rm rad} 
= -\rho \kappa_F \frac{L_{\rm rad}}{4\pi r^2c}.
\end{equation}
This equation can be integrated if an initial value
$ P_{\rm ref}\equiv P_{\rm rad}(R_{\rm ref})$ at a given reference
radius $R_{\rm ref}$, the flux-mean opacity $\kappa_F(r)$, and the
radiative luminosity $L_{\rm rad}(r)$ are known. Our final models in
Sect.\,\ref{sec:numerics} are based on a direct numerical integration
of this equation, taking into account that $L_{\rm rad}$ is a function
of radius (cf.\ Eq.\,\ref{eq:Lrad}).

To describe the solution of Eq.\,\ref{eq:PRGR} analytically,
  we make the simplifying assumption that $L_{\rm rad}=L=const.$
  throughout the wind. As demonstrated by \citet{gra2:13}, integrating
  Eq.\,\ref{eq:PRGR} from $R_{\rm ref}$ to the sonic radius
  $R_{\rm s}$ leads to the expression
\begin{equation}
\label{eq:prad}
P_{\rm rad} = P_{\rm ref} \times \left( \frac{1}{2} + \frac{3}{4} \frac{R_{\rm ref}^2}{R_{\rm s}^2}\bar{\tau}_{\rm s}\right)
\end{equation}
for the radiation pressure at the sonic radius, where $P_{\rm ref}$ is
the radiation pressure at the reference radius $R_{\rm ref}$, and
$\bar{\tau}_{\rm s}$ a weighted mean optical depth of the form
\begin{equation}
  \label{eq:taubar}
  \bar{\tau}_{\rm s} = \int_{R_{\rm s}}^\infty \kappa_F \rho
  \frac{R_{\rm s}^2}{r^2}\, {\rm d}r.
\end{equation}
This expression for $\bar{\tau}_{\rm s}$ is designed to be comparable
to the classical flux-mean optical depth $\tau_{\rm s}$
(cf.\,Eq.\,\ref{eq:motion2}).  Because the integral in
Eq.\,\ref{eq:taubar} is weighted towards small radii with
$r\gtrsim R_{\rm s}$, we have
$\bar{\tau}_{\rm s} \lesssim \tau_{\rm s}$.

To compute $P_{\rm ref}$ and $R_{\rm ref}$ we follow the approach of
\citet{luc1:93,luc1:76} and assume that $P_{\rm ref}$ is connected to
the radiative flux $F_{\rm ref}$ at $R_{\rm ref}$ via an effective
(flux) temperature $T_{\rm ref}$, i.e.,
\begin{equation}
\label{eq:pref}
P_{\rm ref} = \frac{4\sigma}{3c}T_{\rm ref}^4 
= \frac{4}{3c}F_{\rm ref}=\frac{(L/c)}{3\pi R_{\rm ref}^2},
\end{equation}
where $R_{\rm ref}$ is the radius where
$\tilde{\tau}=\bar{\tau}\times(R_{\rm ref}^2/R_{\rm s}^2)=2/3$.  This
relation is to some extent analogous to the concept of the effective
temperature in plane-parallel atmospheres.

With Eq.\,\ref{eq:pref}, Eq.\,\ref{eq:prad} becomes
\begin{equation}
\label{eq:prad2}
  P_{\rm rad}
  =
  \frac{L}{4\pi c} \times 
  \left( \frac{2}{3 R_{\rm ref}^2}
    +\frac{\bar{\tau}_{\rm s}}{ R_{\rm s}^2} \right).
\end{equation}

We note that the only approximations in Eq.\,\ref{eq:prad2} are our
assumptions that $L=L_{\rm rad}$, and that $P_{\rm ref}$ is given by
Eq.\,\ref{eq:pref}. For WR stars $L=L_{\rm rad}$ is usually fulfilled
within a margin of several percent, up to $\approx 10 \%$ for the most
extreme objects with the strongest winds \citep[e.g.][]{cro1:02}. The
validity of Eq.\,\ref{eq:pref} has been confirmed with a high level of
accuracy by \citet{gra2:13}, through direct comparison with a
dynamical non-LTE atmosphere model for an early-type WC star from
\citet{gra1:05}, where the temperature structure is computed from the
requirement of radiative equilibrium \citep[cf.][]{gra1:02,ham1:03}.

In the limit of large wind optical depth we have
$\bar{\tau}_{\rm s} \gg 1$ and $R_{\rm ref} \gg R_{\rm s}$, so that
Eq.\,\ref{eq:prad2} simplifies to
\begin{equation}
  P_{\rm rad}
  = \frac{L}{4\pi R_{\rm s}^2 c} \times \bar{\tau}_{\rm s}
  \equiv \phi_{\rm rad} (R_{\rm s}) \times {\bar{\tau}_{\rm s}},
\end{equation}
where $\phi_{\rm rad}(R_{\rm s})$ denotes the radiative momentum flux
at the sonic radius. Similarly, Eq.\,\ref{eq:pgas} can be expressed as
\begin{equation}
  P_{\rm gas}= \frac{\dot{M}a}{4\pi R_{\rm s}^2}
  \equiv \phi_{\rm wind} (R_{\rm s}),
\end{equation}
where $\phi_{\rm wind}(R_{\rm s})$ denotes the mechanical wind
momentum flux at the sonic point.  The ratio $P_{\rm rad}/P_{\rm gas}$
at the sonic radius can thus be expressed as
\begin{equation}
\label{eq:prpg1}
  \frac{P_{\rm rad}}{P_{\rm gas}}
  = \frac{\phi_{\rm rad}(R_{\rm s})}{\phi_{\rm wind}(R_{\rm s})} 
  \times {\bar{\tau}_{\rm s}}
  \equiv \frac{\Phi_{\rm rad}(R_{\rm s})}{\Phi_{\rm wind}(R_{\rm s})} 
  \times {\bar{\tau}_{\rm s}},
\end{equation}
where $\Phi_{\rm rad}(R_{\rm s})=L/c$ and
$\Phi_{\rm wind}(R_{\rm s})=\dot{M}a$ denote the total radiative and
wind momentum at $R_{\rm s}$.

\subsubsection{Approximate relation for $P_{\rm rad}/P_{\rm gas}$ at
  the sonic point}
\label{sec:eta}

An analytical expression for the wind condition in
Sect.\,\ref{sec:prpg} can be derived based on Eq.\,\ref{eq:prpg1},
using a well-known relation between the wind-efficiency factor
$\eta=\dot{M}\varv_\infty/(L/c)$ (the ratio between the wind momentum
and the momentum of the radiation field) and the sonic-point optical
depth $\tau_{\rm s}$, namely that $\eta \approx \tau_{\rm s}$
\citep[cf.][]{net1:93,gay2:95,lam1:99,vin1:12}.

This relation can be understood under the assumption that photons are
scattered multiple times while escaping the stellar wind, undergoing a
random walk.  For a random walk the effective number of scatterings
that a typical photon must undergo to transfer the necessary momentum
to the wind is $\eta^2$. From the viewpoint of radiative transfer the
number of scatterings in such a picture would be $\tau^2$. As a
consequence $\eta \approx \tau$. Because the layers below the sonic
point are near hydrostatic equilibrium, the relevant optical depth for
the wind acceleration above this point is the sonic-point optical
depth $\tau_{\rm s}$.

To derive the resulting wind condition we start with the equation of
motion for a spherically-expanding, radiatively-driven wind with a
velocity field $\varv(r)$. Above the sonic point, this equation is
well described by
\begin{equation}
  \label{eq:motion1}
  \varv \frac{{\rm d}\varv}{{\rm d}r} = \frac{\kappa_F L}{4\pi r^2 c} - \frac{GM}{r^2},
\end{equation}
where gas-pressure terms are neglected. In this equation the wind
acceleration $\varv \times \varv'$ on the left-hand side, is balanced
by the outward-directed radiative acceleration and the inward-directed
gravitational attraction on the right-hand side. With the Eddington
factor $\Gamma=\kappa_F L_{\rm rad} / (4\pi c GM)$, defined as the
ratio between radiative and gravitational acceleration, this equation
can be rewritten as
\begin{equation}
\label{eq:motion}
  \varv \,{\rm d}\varv = (\Gamma -1) \frac{GM}{r^2}\,{\rm d}r.
\end{equation}

When we integrate this equation from the sonic point (with
$r=R_{\rm s}$, $\varv=a_{\rm s}$) towards infinity (i.e.,
$r\rightarrow \infty$, $\varv\rightarrow\varv_\infty$) the value of
$\Gamma$ will vary as a function of $r$, starting with
$\Gamma \simeq 1$ at $r=R_{\rm s}$, and reaching a maximum near the
point of the largest wind acceleration ${\rm d}\varv/{\rm d}r$. The
value of $\Gamma$ in this region will dominate the integral. For the
qualitative considerations in the remainder of this section we thus
adopt $\Gamma = \Gamma_{\rm w} = const.$ where $\Gamma_{\rm w}$ is
meant to be representative for the value of $\Gamma$ in the region of
highest wind acceleration.  Integration of Eq.\,\ref{eq:motion} then
results in
\begin{equation}
\varv_\infty^2 - a_{\rm s}^2 = (\Gamma_{\rm w} -1)\,\varv_{\rm esc}^2,
\end{equation}
with the escape velocity defined as
$\varv_{\rm esc}=\sqrt{2GM/R_{\rm s}}$. With $\varv_\infty \gg a_{\rm s}$ this
simplifies to
\begin{equation}
\label{eq:VinfVesc}
  \varv_\infty^2 \approx (\Gamma_{\rm w} -1)\,\varv_{\rm esc}^2,
\end{equation}
or equivalently,
\begin{equation}
\label{eq:vfac}
\frac{\Gamma_{\rm w}-1}{\Gamma_{\rm w}}\approx
\frac{\varv_\infty^2}{\varv_\infty^2+\varv_{\rm esc}^2}
= \frac{1}{1+\frac{\varv_{\rm esc}^2}{\varv_\infty^2}}.
\end{equation}
In a second step, Eq.\,\ref{eq:motion} can be rewritten using the
equation of continuity $\dot{M}=4\pi\rho\varv r^2$
\begin{equation}
\label{eq:motion2}
  \dot{M}\,{\rm d}\varv = 4\pi GM\rho\,(\Gamma-1)\,{\rm d}r
 = - \frac{L}{c}\,\frac{\Gamma-1}{\Gamma}\,{\rm d}\tau,
\end{equation}
with the flux-mean optical depth
${\rm d}\tau = - \kappa_F \rho\, {\rm d}r$. Again, integrating from
the sonic point towards infinity, and adopting
$\Gamma=\Gamma_{\rm w}=const.$ we obtain
\begin{equation}
  \dot{M}\,(\varv_\infty -a_{\rm s}) 
= \frac{L}{c}\,\frac{\Gamma_{\rm w}-1}{\Gamma_{\rm w}} \tau_{\rm s}
\end{equation}
with the sonic-point optical depth $\tau_{\rm s}$.  Finally, we obtain
\begin{equation}
\label{eq:eta}
  \eta \equiv \frac{\Phi_{\rm wind}(\infty)}{\Phi_{\rm rad}(\infty)}
  = \frac{\dot{M} \varv_\infty}{L/c} 
  \approx \frac{\Gamma_{\rm w}-1}{\Gamma_{\rm w}} \tau_{\rm s}
  \approx \frac{\tau_{\rm s}}{1+\frac{\varv_{\rm esc}^2}{\varv_\infty^2}}
\end{equation}
for the wind efficiency factor $\eta$, where we made use of
Eq.\,\ref{eq:vfac} and $\varv_\infty \gg a_{\rm s}$.  Analogous to
Eq.\,\ref{eq:prpg1} $\Phi_{\rm rad}(\infty)=L/c$ and
$\Phi_{\rm wind}(\infty)=\dot{M}\varv_\infty$ are the total radiative
and wind momenta for $r \rightarrow \infty$.

Given that usually $\varv_\infty > \varv_{\rm esc}$ for hot-star
winds, this means that the wind efficiency $\eta$ is of the same order
of magnitude as $\tau_{\rm s}$ with a correction factor
$1/(1+\varv_{\rm esc}^2/\varv_\infty^2)$ which reflects the fraction
of the wind momentum that is used to overcome the gravitational
attraction of the star. We note that, primarily due to our assumption
that $\Gamma = \Gamma_{\rm w} = const.$, this relation is only
approximate, but will likely provide a good qualitative picture of the
involved dependencies.

Combining Eqs.\,\ref{eq:prpg1} and \ref{eq:eta} with
$\bar{\tau}_{\rm s} \lesssim \tau_{\rm s}$ (cf.\ Eq.\,\ref{eq:taubar})
we finally obtain
\begin{equation}
  \frac{P_{\rm rad}}{P_{\rm gas}} 
\approx 
\frac{\Phi_{\rm rad}(R_{\rm s})}{\Phi_{\rm wind}(R_{\rm s})}
\times \frac{\Phi_{\rm wind}(\infty)}{\Phi_{\rm rad}(\infty)}
\times \left(1 + \frac{\varv_{\rm esc}^2}{\varv_\infty^2}\right)
\end{equation}
or
\begin{equation}
  \label{eq:psonic}
  \frac{P_{\rm rad}}{P_{\rm gas}} 
\approx 
\frac{\varv_\infty}{a_{\rm s}} 
\times \left(1 + \frac{\varv_{\rm esc}^2}{\varv_\infty^2}\right)
\end{equation}
for the ratio $P_{\rm rad}/P_{\rm gas}$ at the sonic point.  This
means that the ratio of $P_{\rm rad}/P_{\rm gas}$ is of the same order
of magnitude as $\varv_\infty/a_{\rm s}$ and thus almost a constant, in line
with the numerical results of \citet{gra2:13}.

Eq.\,\ref{eq:psonic} is independent of $\dot{M}$, and can be used to
estimate $\dot{M}$ as a function of $\varv_\infty$. To this purpose we
express the sonic speed $a_{\rm s}$ through the radiation pressure
$P_{\rm rad}$ via
\begin{equation}
  a_{\rm s}^8 = \left(\frac{{\cal R} T_{\rm s}}{\mu}\right)^4
  =  \left(\frac{{\cal R}}{\mu}\right)^4 \frac{3c}{4\sigma} \times P_{\rm rad},
\end{equation}
so that Eq.\,\ref{eq:psonic} can be written as
\begin{equation}
\label{eq:psonic2}
P_{\rm gas}
\approx \frac{a_{\rm s}}{\varv_\infty + \frac{\varv_{\rm esc}^2}{\varv_\infty}} 
\times P_{\rm rad}
=  \left(\frac{{\cal R}}{\mu}\right)^{\frac{1}{2}}
\left(\frac{3c}{4\sigma}\right)^{\frac{1}{8}}
\frac{P_{\rm rad}^{\frac{9}{8}}}{\varv_\infty + \frac{\varv_{\rm esc}^2}{\varv_\infty}},
\end{equation}
and with $\mu=4/3$
\begin{equation}
\label{eq:psonic3}
P_{\rm gas}
\approx \frac{5.275 \kms}{\varv_\infty + \frac{\varv_{\rm esc}^2}{\varv_\infty}}
\times 
\frac{P_{\rm rad}^{\frac{9}{8}}}{\rm \left(dyn\,cm^{-2}\right)^{\frac{1}{8}}}.
\end{equation}
This relation provides an estimate of the sonic-point conditions that
are imposed by an optically-thick wind with a specified terminal wind
velocity $\varv_\infty$ in the $P_{\rm rad}$--$P_{\rm gas}$
plane. $\varv_{\rm esc}$ and $\mu$ are given by the stellar mass,
radius, and chemical composition.

\begin{figure}[tbp!]
  \parbox[b]{0.49\textwidth}{\center{\includegraphics[scale=0.35]{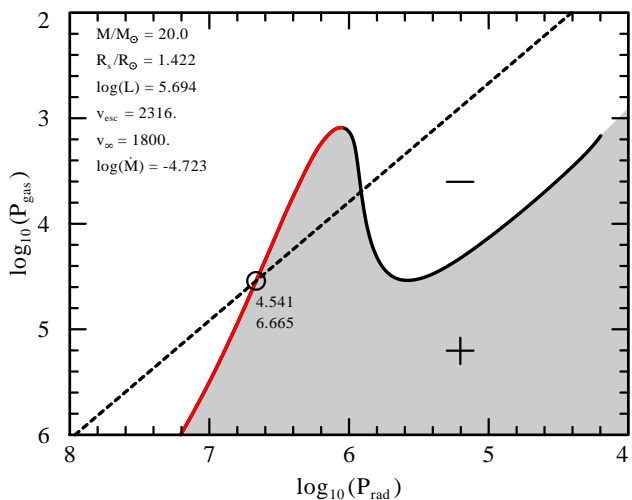}}}
  \caption{{\remark Mass-loss determination in $P_{\rm
        rad}$--$P_{\rm
        gas}$
      plane.} The plot is analogous to Fig.\,\ref{fig:CRIT}, but now
    the black dashed line indicates the sonic-point conditions of an
    optically-thick wind {\remark with an adopted terminal wind
      velocity of $\varv_\infty=1800$\,km/s} following
    Eq.\,\ref{eq:psonic2}. Again, the circle indicates the innermost
    intersection point of Eq.\,\ref{eq:psonic} with the static
    envelope solution. Given the values of $P_{\rm rad}$ and
    $P_{\rm gas}$ at the intersection point, $\dot{M}$ can be
    determined from Eq.\,\ref{eq:mdot}.}
  \label{fig:APPROX}
\end{figure}

The combination of the wind condition (Eq.\,\ref{eq:psonic3}) with the
critical condition $\Gamma \simeq 1$ at the sonic point
(Eq.\,\ref{eq:crit}) provides an estimate of the mass-loss rate of an
optically-thick wind. This is illustrated in Fig.\,\ref{fig:APPROX},
where the intersection point between Eq.\,\ref{eq:psonic3} and the
hydrostatic envelope solution (with $\Gamma \simeq 1$) specifies the
values of $P_{\rm rad}$ and $P_{\rm gas}$ for which envelope and wind
can be connected at the sonic point. The mass-loss rate that
corresponds to these values can then be computed from
Eq.\,\ref{eq:mdot}, in the same way as illustrated in
Fig.\,\ref{fig:CRIT}. For the example in Fig.\,\ref{fig:APPROX} we
obtain a value of $\dot{M}=10^{-4.72}\msunpyr$ for an adopted terminal
wind velocity of $\varv_\infty = 1800\kms$, which lies within the
observed range of WR mass-loss rates.

Notably, the denominator in the last term of Eqs.\,\ref{eq:psonic2}
and \ref{eq:psonic3} has a minimum for $\varv_\infty=\varv_{\rm esc}$.
This means that $P_{\rm gas}$, and thus also $\dot{M}$, are reaching a
maximum for $\varv_\infty=\varv_{\rm esc}$. If the approximations made
in this section were strictly valid, this maximum value would
introduce new upper limits for the mass-loss rates of WR stars and
LBVs.

It is further important to note that Eqs.\,\ref{eq:psonic} to
\ref{eq:psonic3} do not depend on the uncertain clumping properties of
the wind material, as the underlying Eqs.\,\ref{eq:motion} and
\ref{eq:motion2} only rely on the assumption that the winds are
radiatively driven. In case of an inhomogeneous wind structure,
$\tau_{\rm s}$ would thus represent an integral over an effective mean
opacity, as discussed e.g.\ by \citet{osk1:07,owo1:08,sun1:14}. An
analysis involving the ratio $P_{\rm rad}/P_{\rm gas}$, as the one
performed by \citet{gra2:13} thus provides a new important means to
diagnose the clumping properties of the wind material in deep layers
that are not directly observable.

\subsubsection{Numerical wind models}

\label{sec:numerics}
\begin{figure}[tbp!]
  \parbox[b]{0.49\textwidth}{\center{\includegraphics[scale=0.35]{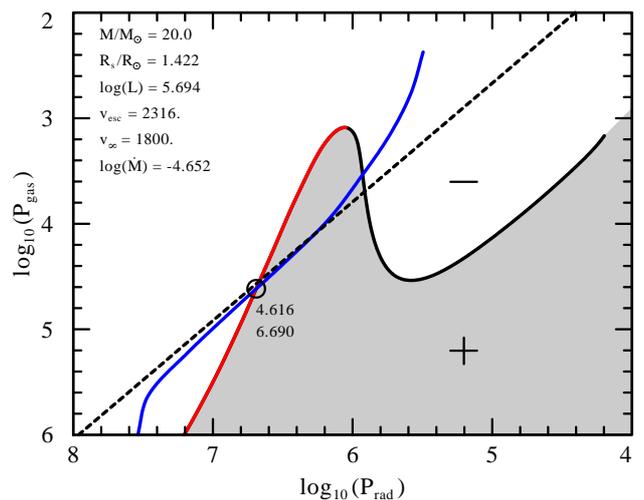}}}
  \caption{{\remark Mass-loss determination in $P_{\rm
        rad}$--$P_{\rm
        gas}$
      plane.} The plot is analogous to Fig.\,\ref{fig:APPROX}, but now
    the sonic-point conditions following from our numerical models
    (indicated by the solid blue curve) are used to determine
    $\dot{M}$.}
  \label{fig:NUMERIC}
\end{figure}

\begin{figure*}[tbp!]
  \parbox[b]{0.99\textwidth}{\center{\includegraphics[scale=0.33]{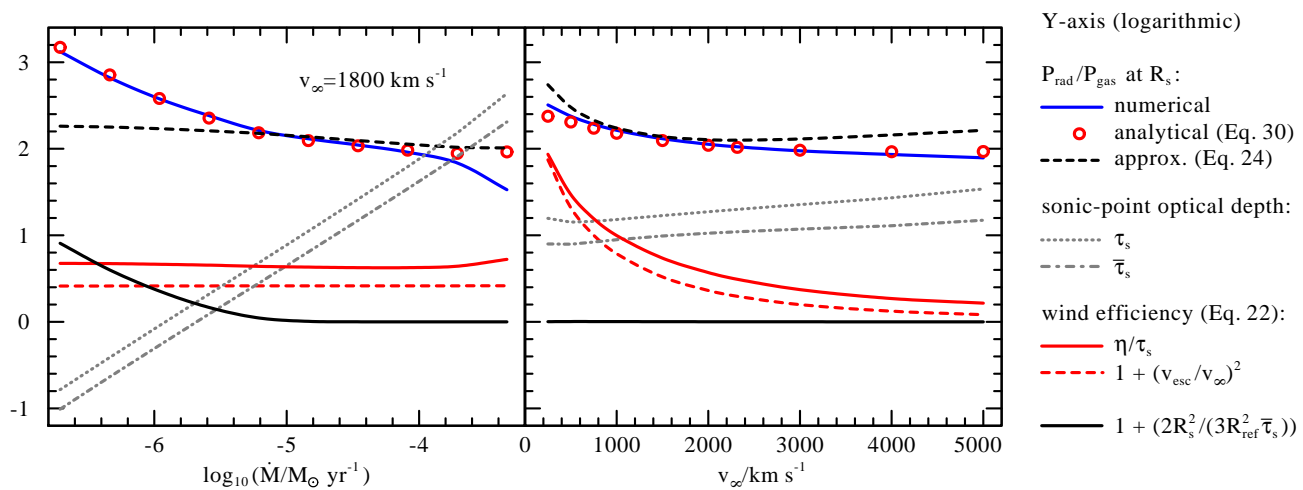}}}
  \caption{{\remark Sonic-point ratio of $P_{\rm rad}/P_{\rm gas}$.}
    {\changed {\remark Different} contributions to the ratio of
      $P_{\rm rad}/P_{\rm gas}$ at the sonic point, as imposed by the
      outer wind, {\remark are indicated} on a logarithmic scale. The
      left panel shows the same sequence of wind models as in
      Fig.\,\ref{fig:NUMERIC}, i.e., models with a fixed terminal wind
      velocity of $\varv_\infty=1800\,\kms$ as a function of the
      adopted mass-loss rate $\dot{M}$. The right panel shows the same
      model sequence as in Fig.\,\ref{fig:MDOT_VINF}, i.e., models for
      which $\dot{M}$ is derived consistently with the critical
      condition (Eq.\,\ref{eq:crit}) as a function {\remark of}
      $\varv_\infty$. Plotted are the ratio of
      $P_{\rm rad}/P_{\rm gas}$ as it follows from our our numerical
      models (solid blue line), from the approximate relation
      Eq.\,\ref{eq:psonic} (black dashed line), and
      Eq.\,\ref{eq:prpg_terms} (red circles).}
    The grey dotted and dash-dotted lines indicate the (weighted)
    sonic-point optical depths $\tau_{\rm s}$ and $\bar{\tau}_{\rm s}$
    respectively. The red solid line indicates the ratio
    $\tau_{\rm s}/\eta$ and the red dashed line the factor
    $(1+\varv_{\rm esc}^2/\varv_\infty^2)$. The black solid line
    indicates the offset factor
    $(1 + {2R_{\rm s}^2}/({3R_{\rm ref}^2\bar{\tau}_{\rm s}}))$
    {\changed in Eq.\,\ref{eq:prpg_terms}}.}
  \label{fig:TERMS_MDOT}
\end{figure*}

In the following we use numerical wind models to derive sonic-point
conditions and mass-loss rates in an analogous way as in
Sect.\,\ref{sec:eta}. To this purpose we follow the approach of
\citet{gra2:13} and solve Eq.\,\ref{eq:PRGR} numerically,
  in line with the weighted mean optical depth $\bar{\tau}_{\rm s}$
  and reference radius $R_{\rm ref}$. To do this we adopt a
  $\beta$-type wind velocity law of the form
\begin{equation}
  \varv(r)=\varv_\infty \times \left( 1 - \frac{R}{r}\right)^\beta
\end{equation}
with $\beta=1$ and a prescribed value of $\varv_\infty$, and compute
$\kappa_{F}(r)$ under the assumption of radiative driving, i.e., in
line with the wind acceleration $\varv \times \varv'$ resulting from
the prescribed velocity law (cf.\ Eq.\,\ref{eq:motion1}). Furthermore,
we compute $L_{\rm rad}(r)$ taking into account that the local
radiative luminosity in a radiatively-driven wind is reduced by the
mechanical wind luminosity, i.e.,
\begin{equation}
  \label{eq:Lrad}
  L_{\rm rad}(r) = L - \dot{M}\times
  \left(\frac{\varv(r)^2}{2} + \frac{MG}{R} - \frac{MG}{r}\right)
  \equiv L - L_{\rm wind}.
\end{equation}
Here we neglect enthalpy terms which turn out to be negligible. The
inclusion of $L_{\rm wind}$ is important for cases where
$L_{\rm wind}\rightarrow L$. In such cases $L_{\rm rad}$ is
significantly reduced by the mechanical work that is needed to
accelerate the wind, and radiative wind driving becomes increasingly
difficult. No wind acceleration is possible beyond the so-called
photon-tiring limit with $L=L_{\rm wind}$ \citep[][cf.\ also
\citealp{heg1:96,gra1:98,owo1:04}]{owo1:97}.

The use of a prescribed velocity law means that we are not solving the
equation of motion explicitly. Instead, we are deriving the
sonic-point conditions {\em under the assumption} that the winds are
radiatively driven. Because the sonic-point values of $P_{\rm rad}$
and $P_{\rm gas}$ are the result of the {\em integrated} radiative
acceleration throughout the outer wind, our results will depend
predominantly on the adopted terminal wind velocity $\varv_\infty$,
and to a lesser extent on the detailed velocity structure.
\citet{gra2:13} performed a detailed comparison of the sonic-point
conditions arising from a $\beta$-type velocity law as we are using it
here, and the dynamically consistent WC star model from
\citet{gra1:05}. Although the model of \citeauthor{gra1:05} displayed
a qualitatively different velocity structure with two acceleration
zones, the resulting sonic-point conditions were remarkably similar,
with differences of 0.06\,dex in $P_{\rm rad}$ and 0.01\,dex in
$P_{\rm gas}$.

The mass-loss rates based on our numerical models are obtained in
analogy to Sect.\,\ref{sec:eta}, i.e., we compute a static stellar
structure model for a given stellar mass $M$, and use the resulting
stellar luminosity $L$ and radius $R$ as an input for our wind
models. As stellar envelopes often inflate near the region of the
Fe-opacity peak, and we are interested in compact solutions where the
sonic point is located just below the inflated region, we do not
extract the formal surface radius from our models, but use the core
radius $R_{\rm c}$ at the bottom of the inflated layer instead
\citep[for more details see][]{gra1:12}. $R_{\rm c}$ serves as an
inner boundary for our wind models, and is usually located very close
to the sonic radius $R_{\rm s}$.

The precise value of $R_{\rm s}$ can only be determined if $\dot{M}$
is known. In our final models in Sect.\,\ref{sec:mdot} we resolve this
problem through an iterative procedure. In the present section we
accept a small inconsistency (up to $\sim 5\%$ in $R$ and $10\%$ in
$\dot{M}$) which arises because $R_{\rm c}$ deviates slightly from
$R_{\rm s}$ in Eq.\,\ref{eq:mdot}.

To derive the sonic-point conditions imposed by the outer wind we
compute a series of wind models (typically for 10 different values of
$\dot{M}$) with the stellar parameters from above, and a given
terminal wind velocity $\varv_\infty$. The resulting sonic-point
values of $P_{\rm rad}$ and $P_{\rm gas}$ are indicated by the blue
curve in Fig.\,\ref{fig:NUMERIC}. The mass-loss rate is computed from
the sonic-point values of $P_{\rm rad}$ and $P_{\rm gas}$ at the
intersection point between the envelope and wind solutions, using
Eq.\,\ref{eq:mdot}. In our example the resulting mass-loss rate of
$\dot{M}= 10^{-4.65}\msunpyr$ compares very well with the approximate
value derived in Sect.\,\ref{sec:eta}.

Fig.\,\ref{fig:NUMERIC} demonstrates that there is a very good
agreement between our approximate wind relation
(Eqs.\,\ref{eq:psonic2}, \ref{eq:psonic3}, indicated by the black
dashed line) and our numerical wind models (blue) in the region near
the actual sonic point. Deviations occur for very high wind densities,
when $L_{\rm wind}$ becomes comparable to $L$, and for low wind
densities, when the wind becomes optically thin.

To better understand the deviations from Eqs.\,\ref{eq:psonic2},
\ref{eq:psonic3} we investigate in Fig.\,\ref{fig:TERMS_MDOT}
{\changed (left panel)} the different contributions to the ratio
$P_{\rm rad}/P_{\rm gas}$ for the model sequence from
Fig.\,\ref{fig:NUMERIC}.  To this purpose we rewrite
Eq.\,\ref{eq:prad2} using the definition of the wind efficiency
$\eta = \dot{M}\varv_\infty/(L/c)$, and divide it by
Eq.\,\ref{eq:pgas} which leads to
\begin{equation}
  \label{eq:prpg_terms}
  \frac{P_{\rm rad}}{P_{\rm gas}}
  =
  \frac{\varv_\infty}{a} \times
  \frac{\tau_{\rm s}}{\eta} \times 
  \frac{\bar{\tau}_{\rm s}}{\tau_{\rm s}} \times 
  \left( 1+\frac{2 R_{\rm s}^2}{3 R_{\rm ref}^2 \bar{\tau}_{\rm s}}\right).
\end{equation}
In Fig.\,\ref{fig:TERMS_MDOT} we extract the different quantities on
the right-hand side (rhs) of this equation from our numerical
models. The approximations that we made when deriving
Eqs.\,\ref{eq:psonic}--\ref{eq:psonic3} were that the second term on
the rhs of Eq.\,\ref{eq:prpg_terms} is well described by
Eq.\,\ref{eq:eta}, and that term three and four are of order unity. We
note that we expect Eq.\,\ref{eq:prpg_terms} to represent our
numerical results very accurately, except for the fact that we assumed
$L_{\rm rad}=const.$ instead of taking Eq.\,\ref{eq:Lrad} into
account.

Our numerical results in Fig.\,\ref{fig:TERMS_MDOT} (indicated by the
blue solid line) deviate from Eq.\,\ref{eq:psonic} (black dashed line)
for low and high $\dot{M}$. For the case of low $\dot{M}$, it is clear
that the optical depths $\tau_{\rm s}$ and $\bar{\tau}_{\rm s}$
(indicated in grey) become too small to justify our assumption of
large optical depth. As a consequence the last term in
Eq.\,\ref{eq:prpg_terms} (black solid line) exceeds unity {\changed
  for low $\dot{M}$}. For high $\dot{M}$ the situation is different,
because the sum of the terms on the rhs of Eq.\,\ref{eq:prpg_terms}
(indicated by red circles) actually displays a very good agreement
with Eq.\,\ref{eq:psonic}. The reason why our numerical results
deviate from this value for high $\dot{M}$, is that the mechanical
wind luminosity $L_{\rm wind}$ becomes comparable to $L$ in
Eq.\,\ref{eq:Lrad}. As a consequence $L_{\rm rad}$ is reduced in the
outer wind, leading to a weaker back-warming effect, and thus lower
$P_{\rm rad}$.

Furthermore, Fig.\,\ref{fig:TERMS_MDOT} demonstrates that the ratio
$\bar{\tau}_{\rm s}/\tau_{\rm s}$ on the rhs of
Eq.\,\ref{eq:prpg_terms} deviates from unity, which is not surprising
given the definition of $\bar{\tau}_{\rm s}$ (cf.\
Eq.\,\ref{eq:taubar} and the related comments). In our example, this
deviation is canceled by the fact that the second term
$\tau_{\rm s}/\eta$ (red solid line) is larger than {\changed what is
  expected from Eq.\,\ref{eq:eta}} and exceeds
$(1+\varv_{\rm esc}^2/\varv_\infty^2)$ (red dashed line) by a similar
amount.

Altogether, our analysis in Fig.\,\ref{fig:TERMS_MDOT} indicates that
Eq.\,\ref{eq:psonic} may indeed be useful to understand the
qualitative dependence of the sonic-point conditions on the outer
wind, but its application in quantitative models is probably hampered
by a too low accuracy. Instead, it is necessary to employ numerical
models for quantitative predictions.

\subsubsection{Mass-loss rates as a function of $\varv_\infty$}
\label{sec:mdot_vinf}

In Sect.\,\ref{sec:numerics} we established a new method to compute
the mass-loss rates of hot stars with optically-thick winds with a
given terminal wind velocity $\varv_\infty$. For practical
applications it will be possible to estimate $\varv_\infty$, e.g.\ for
a certain class of objects, either observationally (based on
spectroscopic results) or from common theoretical arguments (e.g.,
that $\varv_\infty$ should be related to $\varv_{\rm esc}$).  In any
case it will be of primary interest how sensitive our results depend
on the choice of $\varv_\infty$.

\begin{figure}[tbp!]
  \parbox[b]{0.49\textwidth}{\center{\includegraphics[scale=0.36]{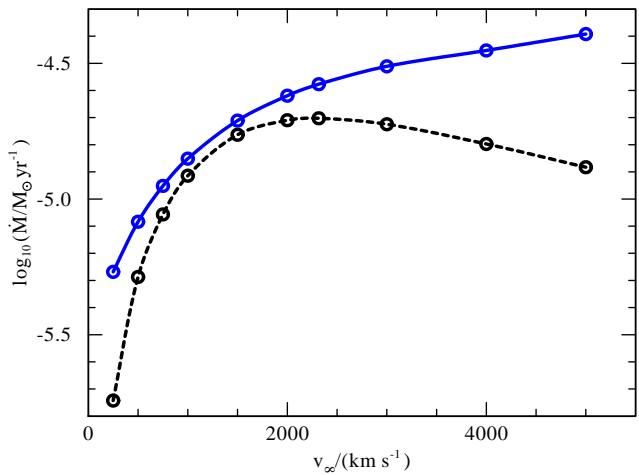}}}
  \caption{{\remark Mass-loss predictions. The derived} mass-loss
    rates for a $20\,M_\odot$ He star {\remark are indicated} as a
    function of the adopted terminal wind velocity $\varv_\infty$. The
    {\changed solid} blue curve indicates numerical wind models
    (cf.\,Sect.\,\ref{sec:numerics}) and the dashed curve our
    approximate estimates (Sect.\,\ref{sec:prpg}).}
  \label{fig:MDOT_VINF}
\end{figure}

For this reason we performed the same computations as described in
Sects.\,\ref{sec:eta} and \ref{sec:numerics} for a $20\,M_\odot$ He
star model for various values of $\varv_\infty$. The results are shown
in Fig.\,\ref{fig:MDOT_VINF}. As expected, the mass-loss rates
obtained from our approximate relation Eq.\,\ref{eq:psonic2} (black
dashed curve in Fig.\,\ref{fig:MDOT_VINF}) show a maximum for
$\varv_\infty = \varv_{\rm esc}$. This maximum does not occur for our
numerical models (blue curve). The reason for this discrepancy is most
likely our simplifying assumption that
$\Gamma = \Gamma_{\rm w} = const.$ in Eqs.\,\ref{eq:vfac} and
\ref{eq:eta}, leading to the given dependence of Eq.\,\ref{eq:psonic}.
Nevertheless, for $\varv_\infty < \varv_{\rm esc}$ the qualitative
behaviour of $\dot{M}$ vs.\ $\varv_\infty$ is well described by our
approximate relation.  Near the maximum of $\dot{M}$ at
$\varv_\infty = \varv_{\rm esc}$ also the numerical results start to
saturate, and the dependence on $\varv_\infty$ becomes relatively
weak.  For the lowest values of $\varv_\infty$ the discrepancy
increases again, most likely because our assumption that
$\varv_{\infty}\gg a$ does not hold anymore.

In {\changed the right panel of Fig.\,\ref{fig:TERMS_MDOT} we show the
  different contributions to the ratio $P_{\rm rad}/P_{\rm gas}$ as a
  function of $\varv_\infty$.} First of all, the figure shows that
$\tau_{\rm s}$ as well as $\bar{\tau}_{\rm s}$ are large enough to
justify our assumption of an optically-thick wind. As a consequence
the last term in Eq.\,\ref{eq:prpg_terms} is indeed very close to
one. As in our example from Sect.\,\ref{sec:numerics} the second and
third term in Eq.\,\ref{eq:prpg_terms} deviate from unity. However, in
the present case the respective errors do not compensate each
other. In particular, the difference between $\tau_{\rm s}$ and
$\bar{\tau}_{\rm s}$ increases for large $\varv_\infty$, leading to a
discrepancy between our approximate and numerical results described
above.

To summarise, it turns out that our approximate relation
Eq.\,\ref{eq:psonic} is not applicable in all cases. However, the
flattening of the relation $\dot{M}(\varv_\infty)$ near the the
maximum at $\varv_\infty = \varv_{\rm esc}$ in Eq.\,\ref{eq:psonic}
also leads to a flattening of this relation in our numerical models,
and thus to a weak dependence of $\dot{M}$ on $\varv_\infty$. This
means that our numerical mass-loss estimates should be reliable, as
long as reasonable estimates of $\varv_\infty$ can be provided.

\section{Mass-loss relations for Wolf-Rayet stars}
\label{sec:galwn}

In this section we use our method from Sect.\,\ref{sec:winds} to
derive mass-loss relations for H-free WN stars. To this purpose we
investigate the terminal wind velocities of these objects in the
Galaxy and the LMC in Sect.\,\ref{sec:vinf}, and compare the resulting
mass-loss relations with empirical results in Sect.\,\ref{sec:mdot}.

\subsection{The terminal wind velocities of H-free WN stars}
\label{sec:vinf}

In this section we investigate the terminal wind velocities
$\varv_\infty$ of H-free WN stars from previous empirical studies of
WR stars in the Galaxy and LMC. The main purpose is to find out which
values of $\varv_\infty$ are appropriate as input parameters for our
models. An important question in this context is how $\varv_\infty$
depends on stellar parameters.

A common way to implement such a dependency is to assume a constant
ratio of $\varv_\infty/\varv_{\rm esc}$. The idea behind this approach
is a (near) equipartition between {\changed the gravitational and
  kinetic contributions to $L_{\rm wind }$ in
  Eq.\,\ref{eq:Lrad}}. Such an equipartition is predicted in the
theory of optically-thin radiatively-driven winds of \citet{cas1:75},
and would also follow from our Eq.\,\ref{eq:VinfVesc} assuming that WR
stars of similar spectral type display similar $\Gamma_{\rm w}$.
Empirically, relations of this form have been confirmed for the winds
of OB stars which tend to show constant ratios of
$\varv_\infty/\varv_{\rm esc}$ well above one \citep[e.g.][]{lam1:95}.
Also for the optically-thick winds of WC stars \citet{gra2:13} found a
relatively clear relation with
$\varv_\infty/\varv_{\rm esc} \approx 1.6$.

\begin{figure}[tbp!]
  \parbox[b]{0.49\textwidth}{\center{\includegraphics[scale=0.44]{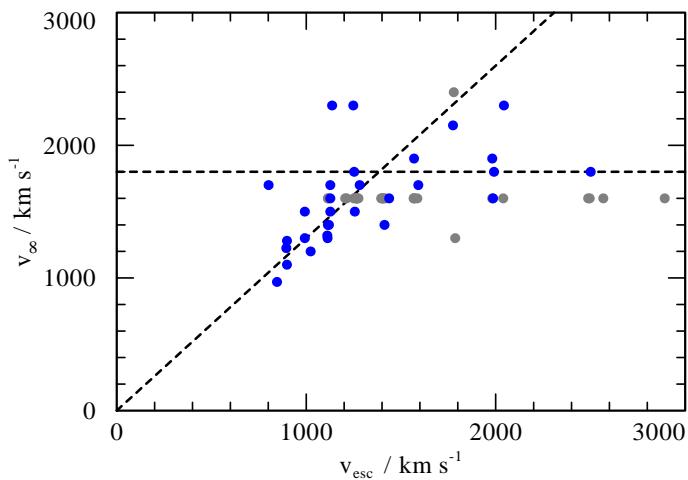}}}
  \caption{{\remark Terminal wind velocities $\varv_\infty$ from
    \citet{ham1:06,hai1:14}. Empirical values for single, H-free WN
    stars in the Galaxy (blue) and LMC (grey) are indicated as a
    function of the escape velocity $\varv_{\rm esc}$.} Dashed lines
  indicate relations with $\varv_\infty/\varv_{\rm esc}=1.3$ and
  $\varv_\infty = 1800$\,km/s.}
  \label{fig:VINF}
\end{figure}

In Fig.\,\ref{fig:VINF} we show the terminal wind velocities
$\varv_\infty$ for putatively single, H-free WN stars in the Galaxy
and LMC as a function of $\varv_{\rm esc}$. The data are compiled from
the comprehensive studies of WN stars in the Galaxy by \citet{ham1:06}
and the LMC by \citet{hai1:14}. While \citeauthor{ham1:06}
concentrated on single WN stars in their work, the study of
\citeauthor{hai1:14} also included binaries and we removed all stars
with indications for multiplicity from the sample. The main parameters
of our sample stars are compiled in Tab.\,\ref{tab:WRGAL} and
\ref{tab:WRLMC}.

\begin{figure}[tbp!]
  \parbox[b]{0.49\textwidth}{\center{\includegraphics[scale=0.44]{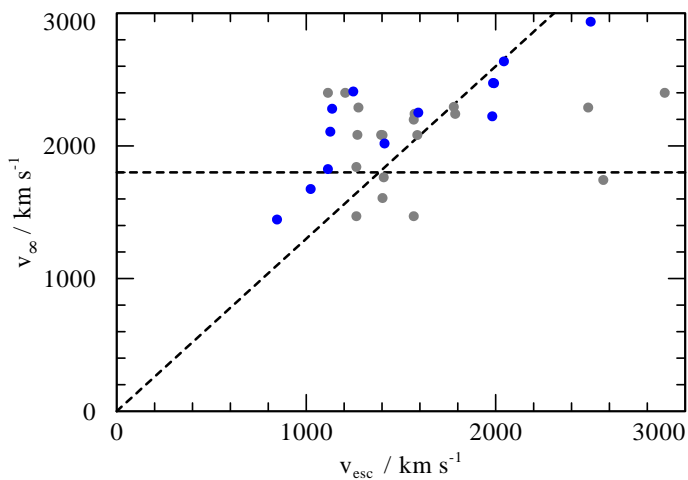}}}
  \caption{{\remark Terminal wind velocities $\varv_\infty$ from
      \citet{nie1:02,nie1:04}. Empirical values for \changed a subset
      of the stars in Fig.\,\ref{fig:VINF} are indicated as a function
      of the escape velocity $\varv_{\rm esc}$.} Dashed lines indicate
    relations with $\varv_\infty/\varv_{\rm esc}=1.3$ and
    $\varv_\infty = 1800$\,km/s.}
  \label{fig:VINF_UV}
\end{figure}

The $\varv_{\rm esc}$ in Fig.\,\ref{fig:VINF} are estimated on the
basis of the spectroscopically determined radii $R_\star$ resulting
from the analyses of \citet{ham1:06,hai1:14}, and stellar masses $M$
obtained from the mass-luminosity relation for pure He-stars from
\citet{gra1:11}. The radii $R_\star$ denote the inner boundary of the
atmosphere models used to analyse the stars.  They are located at
large optical depth (typically of the order of $\tau=20$) in the
nearly hydrostatic layers of the atmosphere models and are almost
identical to the sonic radius $R_{\rm s}$. 

It is important to note that, because of the WR radius problem, the
spectroscopic radii used here do not necessarily resemble the radii
that we compute in our present (compact) stellar structure models
(cf.\ also the discussion in Sect.\,\ref{sec:disc_rv}).  As a
consequence only part of the stars in the observed sample may be
representative for our models. Our present models display escape
velocities in the range $\varv_{\rm esc} \approx 2200$--2600\,km/s,
i.e., they hardly vary.

{\changed In} Fig.\,\ref{fig:VINF} only the coolest stars, with
$\varv_{\rm esc} \lesssim 1200$\,km/s, seem to follow a relation with
$\varv_\infty/\varv_{\rm esc} \approx 1.3$.  At higher escape
velocities this trend vanishes and the terminal wind velocities seem
to saturate. For the Galactic stars this saturation seems to occur
near 1800\,km/s and only few objects lie in the saturated regime.  For
the LMC stars almost {\em all} H-free WN stars seem to lie in the
saturated regime. According to \citet{hai1:14} their spectra are in
agreement with a constant value of $\varv_\infty =1600 \pm 200$\,km/s,
with only two exceptions.

{\changed For comparison, we extracted terminal wind velocities from
  the morphological ananlyses of UV line-profiles of
  \citet{nie1:02,nie1:04}, which are based on spectroscopic data taken
  with the International Ultraviolet Explorer (IUE) sattelite. The
  results for the sub-sample overlapping with our sample are plotted
  in Fig.\,\ref{fig:VINF_UV} (cf.\ also Tab.\,\ref{tab:WRGAL} and
  \ref{tab:WRLMC}). There are significant discrepancies with respect
  to the values derived by \citet{ham1:06,hai1:14}, with the data of
  \citet{nie1:02,nie1:04} tendentially showing higher values of
  $\varv_\infty$. In Fig.\,\ref{fig:VINF_UV} now the complete Galactic
  sample seems to follow a relation with
  $\varv_\infty/\varv_{\rm esc} \approx 1.3$, and the LMC data are,
  with few exceptions, in much better agreement with such a relation.
  These discrepancies have also been discussed by \citet{hai1:14}, who
  noted that adopting the higher $\varv_\infty$ from
  \citeauthor{nie1:04} leads to a mismatch of the line widths in the
  optical range.

  The reason for the discepacies are unclear. Notably, also
  \citet{nie1:04} reported peculiarities and inconsistencies between
  $\varv_\infty$ determined from different ionic species.  In
  particular for the LMC the quality of the IUE data is partly very
  poor. Moreover, the pointing accuracy and large aperture size of IUE
  may not be sufficient to separate stars in dense fields in the
  LMC. One could also speculate that the UV profiles do not always
  stem from the same cool wind component as the optical lines, similar
  to what has been discussed for clumped, multi-component winds of
  O\,stars \citep{sun1:14}.}

In the following analysis we {\changed acknowledge that the terminal
  wind velocities of WR stars are uncertain, and} compute different
model sequences with $\varv_\infty = const.$ and
$\varv_\infty/\varv_{\rm esc} = const.$

\subsection{WR mass-loss rates}
\label{sec:mdot}

In this section we use our models from Sect.\,\ref{sec:numerics} in
combination with terminal wind velocities from Sect.\,\ref{sec:vinf}
to predict mass-loss rates of H-free WN stars. We computed two model
sequences for Galactic metallicity ($Z=0.02$), one with a fixed
terminal wind velocity of $\varv_\infty= 1800\,\kms$, and one with
$\varv_\infty/\varv_{\rm esc} = 1.3$. For the compact stellar
structure models used in this work the latter relation results in
values of $\varv_\infty = 2800$--3400\,\kms. For LMC metallicity
($Z=0.008$) we computed two sequences with $\varv_\infty= 1600\,\kms$
and $\varv_\infty/\varv_{\rm esc} = 1.3$.

\begin{figure}[t!]
  \parbox[b]{0.49\textwidth}{\center{\includegraphics[scale=0.36]{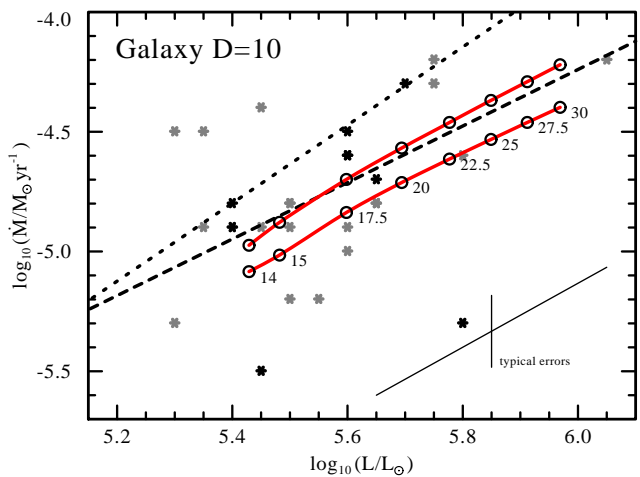}}}
  \caption{{\remark Mass-loss relations for Galactic metallicity.}
    Mass-loss rates for H-free WN stars in the Galaxy as a function of
    luminosity. Black/grey stars indicate empirical mass-loss rates
    for Galactic WN stars with/without known distance from cluster
    membership \citep{ham1:06}. The empirical mass-loss rates have
    been scaled down to those expected for a wind clumping factor of
    $D=10$.  The red curves indicate mass-loss rates obtained from our
    models with $Z=0.02$ for $\varv_\infty/\varv_{\rm esc}=1.3$ (top)
    and $\varv_\infty=1800\,\kms$ (bottom), {\changed the labels
      indicate stellar masses in $M_\odot$}. The dotted/dashed lines
    indicate the empirical mass-loss prescriptions for H-free WN stars
    from \citet{nug1:00} and \citet{yoo1:17}.}
  \label{fig:MDOT_GAL}
\end{figure}

\begin{figure}[t!]
  \parbox[b]{0.49\textwidth}{\center{\includegraphics[scale=0.36]{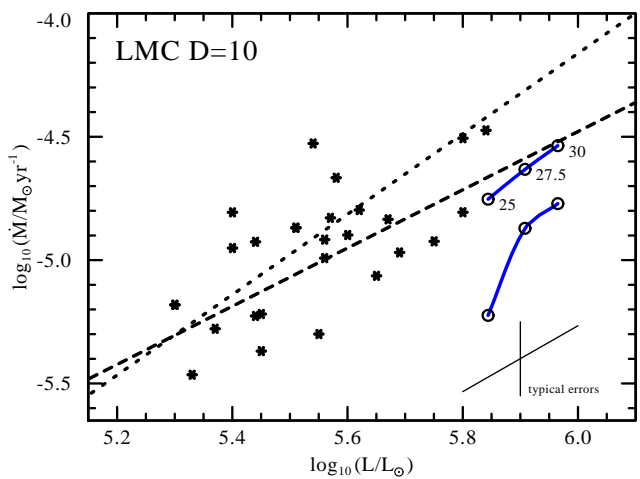}}}
  \caption{{\remark Mass-loss relations for LMC metallicity.}
    Mass-loss rates for H-free WN stars in the LMC as a function of
    luminosity. Black stars indicate empirical mass-loss rates from
    \citep{hai1:14}. The blue curves indicate mass-loss rates obtained
    from our models with $Z=0.008$ for
    $\varv_\infty/\varv_{\rm esc}=1.3$ (top) and
    $\varv_\infty=1600\,\kms$ (bottom), {\changed the labels indicate
      stellar masses in $M_\odot$}. The dotted/dashed lines indicate
    the empirical mass-loss prescriptions for H-free WN stars from
    \citet{nug1:00} scaled with the metallicity dependence from
    \citet{vin1:05}, {\changed and from \citet{yoo1:17} respectively.
      Notably, our models do not predict solutions below
      $\sim 25\,M_\odot$, which is in conflict with the observed range
      of luminosities for H-free WN stars in the LMC.}}
  \label{fig:MDOT_LMC}
\end{figure}

In contrast to our previous computations we estimated the sonic radius
$R_{\rm s}$ as precisely as possible by matching the sonic-point
density following from Eq.\,\ref{eq:pgas} iteratively for each given
value of $\dot{M}$ with the corresponding density in our static
stellar structure models. The resulting mass-loss rates $\dot{M}$ are
shown as a function of luminosity $L$ in Figs.\,\ref{fig:MDOT_GAL} and
\ref{fig:MDOT_LMC}.

In these two figures we compare our results with our samples of
putatively single H-free WN stars in the Galaxy and LMC from
Tab.\,\ref{tab:WRGAL} and \ref{tab:WRLMC} extracted from
\citet{ham1:06,hai1:14}. Because different wind clumping factors $D$
were adopted in these two studies we scaled the mass-loss rates from
\citeauthor{ham1:06} (which were originally obtained with $D=4$) with
$\dot{M}\propto 1/\sqrt{D}$ to match the ones from
\citeauthor{hai1:14} with $D=10$. The same inconsistency has recently
been pointed out by \citet{yoo1:17} who {\changed obtained a
  clumping-corrected relation for H-free WN stars by fitting the
  corrected data from \citet{ham1:06} using the luminosity dependence
  with $\dot{M}\propto L^{1.18}$ derived by \citet{hai1:14} for WN
  stars in the LMC}.

In Figs.\,\ref{fig:MDOT_GAL} and \ref{fig:MDOT_LMC} we compare our
results with the commonly used empirical mass-loss relation from
\citet{nug1:00} scaled with the $Z$-dependence from \citet{vin1:05},
i.e.,
\begin{equation}
  \label{eq:NL2000}
  \log\left(\frac{\dot{M}}{\msunpyr}\right)
  = 1.63 \times \log\left(\frac{L}{L_\odot}\right) 
  + 0.86 \times \log\left(\frac{Z}{Z_\odot}\right) -13.6,
\end{equation}
and the {\changed relation from} \citet{yoo1:17}, i.e.,
\begin{equation}
  \label{eq:YOO2017}
  \log\left(\frac{\dot{M}}{\msunpyr}\right)
  = 1.18 \times \log\left(\frac{L}{L_\odot}\right) 
  + 0.6 \times \log\left(\frac{Z}{Z_\odot}\right) - 11.32.
\end{equation}

The theoretical relations from our Galactic models in
Fig.\,\ref{fig:MDOT_GAL} display a nice agreement with relation from
\citet{yoo1:17}. {\changed In particular} the slope of our predicted
mass-loss relations (with roughly $\dot{M}\propto L^{1.3}$) compares
well with the slope proposed by \citeauthor{yoo1:17}
($\dot{M}\propto L^{1.18}$) and is significantly shallower than the
relation from \citeauthor{nug1:00} ($\dot{M}\propto L^{1.63}$).

\begin{figure}[tbp!]
  \parbox[b]{0.49\textwidth}{\center{\includegraphics[scale=0.35]{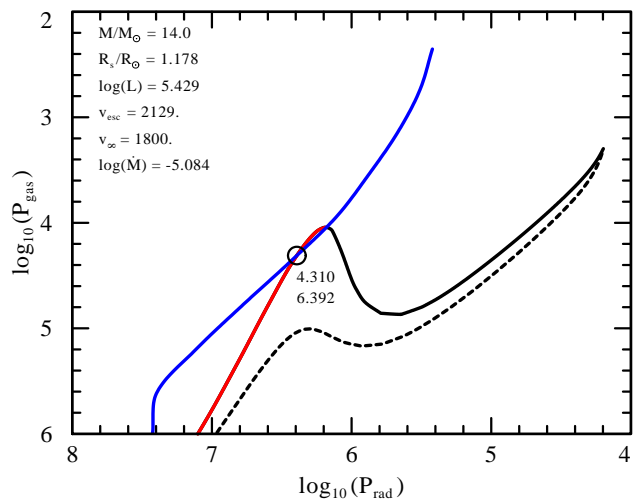}}}
  \caption{Limiting masses. Wind and envelope solutions for a
    $14\,M_\odot$ He star analogous to Fig.\,\ref{fig:NUMERIC}. For
    $Z=0.02$ the envelope solution (red/black curve) just matches the
    wind solution with $\varv_\infty=1800$\,km/s (blue curve). For
    $Z=0.008$ (black dashed curve) no match can be obtained.}
  \label{fig:TEST008}
\end{figure}

Another important result is that we encounter {\changed stellar} mass
limits below which no wind solutions exist. For our Galactic models
this situation occurs for models $\lesssim 14\,M_\odot$. To illustrate
the situation we show the solution topology for our lowest-mass model
with $14\,M_\odot$ in Fig.\,\ref{fig:TEST008}.  For this mass the wind
solution with $\varv_\infty=1800$\,km/s just touches the tip of the
envelope solution, so that there still exists an intersection
point. For lower masses, the $L/M$ ratio {\changed would decrease}, so
that the Eddington opacity $\kappa_{\rm Edd} = 4\pi c G M /L$, which
needs to be matched at the sonic point due to the critical condition
(Eq.\,\ref{eq:crit}), {\changed would increase}.  As a consequence
higher densities {\changed would be} required at the sonic point, and
the envelope solution in Fig.\,\ref{fig:TEST008} {\changed would move}
downward, towards higher values of $P_{\rm gas}$. In such a case
{\changed there would be no intersection point between envelope and
  wind solution, which means that a connection of wind and envelope at
  the sonic point would be physically impossible.}

The same situation occurs for lower metallicities. The envelope
solution for a $14\,M_\odot$ model with LMC metallicity is indicated
as a dashed curve in Fig.\,\ref{fig:TEST008}. Also in this case,
higher densities are needed to produce the required value of
$\kappa_{\rm Edd}$. Clearly, no intersection point exists between the
dashed envelope solution and the wind solution in
Fig.\,\ref{fig:TEST008}.

To shift the wind solution accordingly, it would be necessary to
reduce the ratio of $P_{\rm rad}/P_{\rm gas}$ in our example by a
factor of about five. Since $P_{\rm rad}$ is confined within a small
range to match the temperature of the Fe opacity peak, this would mean
that $P_{\rm gas}$, and thus $\dot{M}$ (according to
Eq.\,\ref{eq:pgas}), would need increase by a similar
factor. Following our discussion in Sect.\,\ref{sec:vinf} such an
increase would demand for a substantial increase of $\varv_\infty$
which is not observed. Because the mechanical part of the wind
luminosity $L_{\rm wind}$ in Eq.\,\ref{eq:Lrad} increases with
$\varv_\infty^2$ such a wind will approach the photon-tiring limit
(with $L = L_{\rm wind}$, cf.\ Eq.\,\ref{eq:Lrad}) before the desired
ratio of $P_{\rm rad}/P_{\rm gas}$ is reached.

As a consequence of the metallicity-dependence of the mass limits
discussed above, our LMC models with $Z=0.008$ in
Fig.\,\ref{fig:MDOT_LMC} only produce solutions for masses
$\gtrsim 25\,M_\odot$. This limit corresponds to a minimum luminosity
of $\log(L/L\odot) \gtrsim 5.8$, a value above which hardly any stars
are observed. Clearly, our models do not reproduce the optically-thick
winds of the bulk of the H-free WN stars in the LMC.

Our high-mass models with $M \gtrsim 25\,M_\odot$, suggest a
metallicity dependence $\dot{M} \propto Z^\gamma$ with an exponent of
$\gamma = 0.8$--1.0, which is comparable to the value of
$\gamma =0.86$ from \citet{vin1:05}, and higher than the value of
$\gamma =0.6$ {\changed which has been derived by \citet{yoo1:17}
  based on observations of H-free WN stars from \citet{ham1:06} and
  the $\dot{M}(L)$ relation for LMC stars from \citet{hai1:14}}.

\section{Discussion}
\label{sec:discussion}

In this work we investigated whether WR-type stellar winds can be
described by the 'classical' assumptions of a compact stellar
structure and a radiatively-driven, optically-thick, smooth stellar
wind. As it became clear in the previous sections, these classical
assumptions can describe many of the observed properties of H-free WN
stars well, but do also fail in several respects. In the following we
summarise and discuss our results. In Sect.\,\ref{sec:disc_winds} we
start with a discussion of the wind physics implied by our models as
well as their possible shortcomings, followed by a discussion of their
dependence on $\varv_\infty$ in Sect.\,\ref{sec:disc_vinf}. In
Sect.\,\ref{sec:disc_rv} we discuss the implications of the observed
terminal wind velocities, and in Sect.\,\ref{sec:disc_mdot} the
relevance of the derived WR mass-loss relations and mass limits. In
Sect.\,\ref{sec:disc_prev} we compare our results with previous
theoretical studies.

\subsection{Wind physics}
 \label{sec:disc_winds}

 In Sect.\,\ref{sec:mdot} we provided the first {\changed theoretical
   $\dot{M}(L)$ relations} for WR stars that are obtained through a
 combination of the critical condition occurring at the sonic point of
 optically-thick winds (Eq.\,\ref{eq:crit}), and the sonic-point
 condition arising from the back-warming effect of the outer wind (the
 'wind condition', Eq.\,\ref{eq:prad2}). While the critical condition
 only depends on the local conditions at the sonic point, the wind
 condition depends on the integrated properties of the outer wind
 above the sonic point, in particular on its optical depth. The
 mass-loss rates obtained from our models are thus a consequence of
 both, the local conditions at the sonic point {\em and} the physics
 of the outer wind.

 The temperature structure resulting from Eq.\,\ref{eq:prad2} is a
 consequence of radiative equilibrium.  The interplay between the
 sonic point and the outer wind is thus moderated by radiative heat
 transfer.  This means that information on the state of the outer wind
 is communicated downstream, towards the sonic point, through
 radiation.

 In our current models we assume that the outer wind is
 radiatively driven. Notably, we do not explain how the wind
 acceleration is sustained, especially beyond the Fe-opacity peak. In
 previous wind models by \citet{gra1:05} this was achieved by non-LTE
 effects, namely through the population of meta-stable energy levels
 that typically occur directly above the ground state of Fe-group
 ions. Nevertheless, \citeauthor{gra1:05} reported that WR wind
 driving is still problematic in the general case, due to a lack of
 opacities in the gap between the Fe-opacity peak and the opacity peak
 at cooler temperatures.

 A possible way to increase the flux-mean opacity in stellar winds is
 the line-deshadowing effect in the presence of velocity gradients. As
 discussed in Sect.\,\ref{sec:winds} the occurrence of this effect is
 governed by the {\changed CAK optocal depth parameter $t_{\rm CAK}$
   (Eq.\,\ref{eq:CAK}).  A critical parameter in Eq.\,\ref{eq:CAK}} is
 the line-broadening velocity $\varv_{\rm Dop}$.  Identifying
 $\varv_{\rm Dop}$ with the thermal velocity of protons
 \citet{nug1:02,ro1:16} argued that Doppler shifts are not important
 in the sonic-point region of WR-type winds. Also in the models of
 \citet{gra1:05} high values of $\varv_{\rm Dop}$ were adopted and
 velocity gradients played a subordinate role.

 Given the extremely narrow thermal width of iron lines, however,
 $\varv_{\rm Dop}$ may be much smaller. This could mean that the
 effects of velocity gradients have been underestimated in previous
 studies. In the presence of high line densities this could be further
 facilitated by multi-line scattering effects \citep{fri1:83,gay2:95},
 potentially leading to a violation of the diffusion approximation
 even for near-sonic velocities.

 In this context it is crucial whether the line-broadening velocity
 $\varv_{\rm Dop}$ is dominated by the slow thermal motions of Fe
 ions, or by turbulent motions which may be orders of magnitude
 higher. While e.g.\ \citet{gra2:16} argued for a turbulent structure
 {\changed in the sub-sonic part} of WR envelopes, the low values of
 $\varv_{\rm Dop}$ may be supported by the suppression of convection
 in the presence of velocity gradients as discussed by
 \citet{ro1:16}. Therefore, at the current stage, we cannot exclude
 that velocity gradients play an important role in WR wind driving, in
 contrast to the underlying model assumptions in this work. In any
 case, our present models serve as an important test case for WR wind
 theories.

\subsection{Dependence on wind velocities}
\label{sec:disc_vinf}

An interesting new aspect of our wind models is their dependence on
the terminal wind velocity $\varv_\infty$, which is used as an input
parameter in our models. {\changed In nature $\varv_\infty$ is set by
  the complex physics of the outer wind. There are, however,
  situations where small changes in stellar parameters lead to drastic
  changes in the outer wind structure and $\varv_\infty$. A well-known
  example is the bi-stability jump for B-supergiants
  \citep{vin1:99,pet1:16}. Also differences in the chemical
  composition will be able to cause variations of
  $\varv_\infty$. Because we are not computing the velocity structure
  self-consistently, such situations can only be considered in our
  models by varying $\varv_\infty$ as input parameter.}

In Sect.\,\ref{sec:mdot_vinf} we found that $\dot{M}$ increases with
increasing $\varv_\infty$ for our optically-thick wind models. This is
in stark contrast to what is found for the optically-thin winds of OB
stars for which the wind momentum $\dot{M}\varv_\infty$ is expected to
be almost invariant for fixed stellar parameters. {\changed As a
  consequence the wind models for OB stars by \citet{vin1:00,vin1:02}
  display the opposite behaviour as ours, namely that $\dot{M}$
  decreases for increasing $\varv_\infty$.}

The reason for the peculiar behaviour of optically-thick winds is
that, due to the effect of the outer wind, the ratio of
$P_{\rm rad}/P_{\rm gas}$ at the sonic point decreases for increasing
values of $\varv_\infty$ (corresponding to the case with
$\varv_\infty < \varv_{\rm esc}$ in Eq.\,\ref{eq:psonic}).  Because
$P_{\rm rad}$ is constrained through the temperature of the Fe-opacity
peak, this means that $P_{\rm gas}$, and thus the density
$\rho_{\rm s}$, at the sonic point increases for increasing
$\varv_\infty$. As a consequence also the mass-loss rate
$\dot{M}=4\pi R^2_{\rm s} \rho_{\rm s} a$ increases for increasing
$\varv_\infty$. This may be the reason why the observed WR wind
velocities saturate easily, leading to terminal wind velocities
$\varv_\infty$ of similar order or even lower than the escape velocity
$\varv_{\rm esc}$.

\subsection{Observed wind velocities and radii}
\label{sec:disc_rv}

The terminal wind velocity is the key parameter for the
parametrisation of the outer wind in our models. For this reason we
reviewed the observed wind properties of H-free WN stars in
Sect.\,\ref{sec:vinf} {\changed using stellar parameters and
  $\varv_\infty$ from \citet{ham1:06,hai1:14}, and alternative values
  $\varv_\infty^{\rm\sc uv}$ determined from UV line profiles by
  \citet{nie1:02,nie1:04}.}

Based on the spectroscopically determined WR radii {\changed and
  $\varv_\infty$ from \citet{ham1:06,hai1:14}}, we found that
predominantly late spectral sub-types in the Galaxy seem to follow a
relation with $\varv_\infty/\varv_{\rm esc} = const.$, similar to the
relation for WC stars from \citet{gra2:13}. The rest of our sample
stars seems to display almost constant values of
$\varv_\infty$. {\changed Using the $\varv_\infty^{\rm\sc uv}$ from
  \citet{nie1:02,nie1:04}, all Galactic objects and, with few
  exeptions, also the LMC objects seem to follow a relation with
  $\varv_\infty/\varv_{\rm esc} = const.$ }

The existence of such relations is relevant for the WR radius problem,
i.e., for the question whether the radii of WR stars are compact, as
adopted in our present models, or if they are much larger, as
suggested by spectroscopic analyses.  Whether or not the spectroscopic
radii are reliable depends on the wind density. \citet{ham1:04}
discussed the limit where the visible spectrum is formed so far out in
the wind that it is not possible to determine the radius of the wind
base. \citet{san1:12,hai1:14} discussed that this limit predominantly
affects the earliest WR sub-types, and that the spectroscopic radii of
most late-type WR stars are most likely well determined.

Based on our discussion in Sect.\,\ref{sec:vinf} the existence of a
relation with $\varv_\infty/\varv_{\rm esc} = const.$ for {\changed WC
  stars and at least part of the WN stars} suggests that their wind
base may indeed be located at larger radii. {\changed On the other
  hand,} a constant value of $\varv_\infty$ {\changed would suggest}
that also $\varv_{\rm esc}$ is (almost) constant, as it is expected
for compact WR stars. {\changed Because we cannot clearly distinguish
  between both scenarios based on the current data, it is} currently
not clear whether WR stars are generally compact or not. However, it
is important to keep in mind that our assumption of a compact core may
not be correct for all stars in the sample.

For our final model computations we used both approaches, i.e., we
used prescriptions with $\varv_\infty/\varv_{\rm esc} = const.$ and
$\varv_\infty = const$.  Because the first relation applies to late
sub-types with large radii, the resulting values for our compact
models ($\varv_\infty = 2800$--3400\,km/s) are larger than what is
typically observed for this type of objects. Nevertheless, it is
important to investigate how different assumptions on $\varv_\infty$
affect our results.

\subsection{Mass-loss rates and mass limits}
\label{sec:disc_mdot}

In Sect.\,\ref{sec:mdot} we compared the $\dot{M}(L)$ relations
resulting from our models with observations. We found a good agreement
with the new empirical mass-loss relation for H-free WN stars of
\citet{yoo1:17}. Empirical mass-loss relations for WR stars are
commonly characterised by power-laws with fixed exponents for the
dependence on luminosity and metallicity. Our models display such a
relation with $\dot{M}\propto L^\delta\, Z^\gamma$, $\delta=1.3$, and
$\gamma=0.8$--1.0. The exponents in this relation do not depend
significantly on the adopted prescription for $\varv_\infty$.

In an absolute sense our predictions are in good agreement with the
mass-loss prescription of \citet{yoo1:17}. However, the results depend
upon the adopted prescription for $\varv_\infty$, in the sense that
the models with constant (and lower) $\varv_\infty$ predict slightly
lower mass-loss rates (cf.\,Sect.\,\ref{sec:disc_vinf}).

The most notable difference between the observed WR population and our
models are the $Z$-dependent mass limits below which our models do not
provide wind solutions. In particular for the low metallicity in the
LMC we find that the Fe peak does not provide enough opacity to
support WR-type mass-loss rates in the region around the sonic-point.

A possible solution to this problem has been discussed previously by
\citet{gra2:13}. These authors suggested that clumping could increase
the mean opacity at the sonic point. According to \citet{gra1:12}
clumping could also lead to an enhanced inflation effect, and would
thus help to explain the large radii that are derived in spectral
analyses of WR stars.  However, as we discussed earlier such a
configuration cannot be realised with a smooth and continuous wind
flow \citep[cf.][]{ro1:16} and would most likely require a very
complex wind structure. An alternative way to increase the flux-mean
opacity near the sonic point could be the line-deshadowing and
multi-line scattering effects discussed in
Sect.\,\ref{sec:disc_winds}.

In this context it is interesting that, according to \citet{san1:12},
the population of WC stars in the Galaxy shows a clear dichotomy
between early and late WC sub-types, where the late sub-types occur
predominantly at low luminosities $\lesssim 10^{5.3}\,L_\odot$.
\citeauthor{san1:12} argue that the large radii inferred for the
latter group suggest that they are inflated. According to our models
{\changed there are indeed no continuous wind solutions for}
smooth/compact WR-type winds in this luminosity regime, so a
clumped/inflated wind structure may be preferred.

{\changed As discussed in Sect.\,\ref{sec:mdot} the encountered mass
  limits result from the critical condition (Eq.\,\ref{eq:crit}) that
  the Eddington opacity $\kappa_{\rm Edd}=4\pi cGM/L$ needs to be
  matched at the sonic point. For a given chemical surface
  composition, the truly decisive parameter is thus the $L/M$ ratio
  rather than the stellar mass. Only for objects with a given $L(M)$
  relation, as it is the case for the core He-burning objects
  investigated here, the encountered limits translate into limits for
  the stellar mass $M$, or alternatively for the luminosity $L$. In
  particular, the derived mass limits do not apply to [WR]-type
  central stars of PNe which are in a phase of shell burning. Strictly
  speaking, our results only apply to pure He stars. In reality the
  cores of He-burning stars become enriched in C and O during their
  evolution. This leads to a slight increase in mean molecular weight,
  and consequently, to an increase of the $L/M$ ratio. However, this
  effect is not strong enough to resolve the discrepancies that we
  encounter at LMC metallicity.}

Even if the encountered mass limits need to be modified, the existence
of physical mass limits {\em per se} is interesting. In fact, the WR
stars in the Galaxy and LMC, including the samples discussed here, do
display lower luminosity/mass limits. In the framework of single-star
evolution such mass limits are expected {\changed (although} current
evolution models struggle to reproduce the luminosities of the
observed WR population \citep[e.g.][]{san1:12}). On the other hand,
binary-interaction models predict the existence of a continuous mass
distribution of WR stars {\changed in binary systems, including
  low-mass WR stars which} are believed to be SN\,Ib/c progenitors
\citep[e.g.][]{smi1:17,zap1:17}.  In this framework the existence of
mass limits for optically-thick winds could explain why {\changed such
  low-mass WR stars have not yet been observed}.  He-stars below the
corresponding WR luminosity limits would not be able to support
optically-thick winds, i.e., they would not display the prototypical
WR emission lines.

\subsection{Comparison with previous wind models}
\label{sec:disc_prev}

Previous models for core He-burning WR stars have been computed based
on Monte-Carlo models \citep{luc1:93,vin1:05}, and based on
hydrodynamic non-LTE model atmospheres \citep{gra1:05}.

The Monte-Carlo models focused {\em only} on the outer wind using
$\beta$-type velocity laws with prescribed $\varv_\infty$. Mass-loss
rates $\dot{M}$ were determined via the condition that the global wind
luminosity (Eq.\,\ref{eq:Lrad}) equals the work performed by the
radiative force resulting from the Mote-Carlo computations. This means
that no local dynamical consistency was obtained, and in particular
that the critical and wind conditions at the sonic point were
ignored. Our present models are complementary to this approach in the
sense that we {\em assume} that the outer wind is radiatively driven
and that we focus only on the resulting conditions at the sonic point.
In comparison with our $\dot{M}(L)$ relations in
Fig.\,\ref{fig:MDOT_GAL} the Galactic early-type WN model of
\citeauthor{luc1:93} with
$[\log(L/L_\odot), \log(\dot{M}/\msunpyr)]=[5.45,-4.68]$ lies
0.2--0.3\,dex above our results. The late-type WN and WC models of
\citeauthor{vin1:05} with $[5.62,-4.89]$ and $[5.36,-5.34]$ lie
$\approx$\,0.1\,dex and 0.2--0.3\,dex below our relation. However, for
these models much larger radii were adopted and the WN model had a
non-zero hydrogen surface mass fraction of $X=0.15$.

The WC model by \citet{gra1:05} was obtained by a simultaneous and
locally consistent numerical solution of the wind hydrodynamics and
the non-LTE radiative transfer in the co-moving frame. This means that
the velocity structure, and in particular $\varv_\infty$, were
computed within the model, and that the critical-point and wind
conditions were fulfilled in the same way as in our present
models. Furthermore, the model by \citet{gra1:05} is equivalent to the
models presented here, in the sense that it has a smooth/compact wind
structure with a sonic point on the hot side of the Fe-opacity
peak. With $[\log(L/L_\odot), \log(\dot{M}/\msunpyr)]=[5.45,-5.14]$
the WC model lies near the lower mass limit obtained in the present
work, roughly 0.1\,dex below the $\dot{M}(L)$ relations for WN stars
in Fig.\,\ref{fig:MDOT_GAL}.

Because of the similarity to our present models, the limitations found
in the present work will also apply to the models by
\citet{gra1:05}. This explains why it was not possible for
\citeauthor{gra1:05} to provide models for metallicities below solar,
while \citet{vin1:05} could provide wind models down to
$\log(Z/Z_\odot) = -4.5$ without problems.

\section{Conclusions}

In this work we examined the physics of optically-thick,
radiatively-driven winds with a smooth wind structure near the sonic
point. We focused on H-free WN stars with a compact structure (without
envelope inflation) and a high sonic-point temperature. For solar
metallicity our models provide $\dot{M}(L)$ relations in good
agreement with observations. In particular their slope agrees well
with commonly used mass-loss prescriptions. However, we encounter mass
limits that suggest that smooth/compact wind models cannot explain the
occurrence of WR-type winds at low masses and/or low metallicities. As
suggested by \citet{gra2:13}, this could mean that the winds of many
WR stars are clumped near the sonic point. Alternatively the Fe-peak
opacities could be enhanced by multi-line effects, or other mechanisms
than radiative driving could play a role in WR wind driving.

The presence of physical mass limits for WR-type winds might play a
role for the non-detection of low-mass WR stars. These possible
SN\,Ib/c progenitors are believed to be formed through binary
interaction but have not yet been detected spectroscopically.  In the
absence of strong stellar winds such objects would not display the
prototypical WR features, and would become very difficult to detect.

The new sonic-point conditions derived in this work suggest that the
ratio $P_{\rm rad}/P_{\rm gas}$ at the sonic radius of optically-thick
winds is of the order of $\varv_\infty/a_{\rm s}$. Because usually
$\varv_\infty \gg a_{\rm s}$, this means that the formation of
optically-thick winds is only possible when radiation pressure
dominates over gas pressure near the sonic radius. This implies that
high Eddington factors are a pre-requisite for the formation of
optically-thick winds.

Otherwise our models show wind properties that
are different from those of OB stars, in particular they show a
qualitatively different dependence on the terminal wind velocity.

The sonic-point conditions derived in the present work can serve as a
new means to connect stellar structure models and optically-thick
winds in a consistent manner at the sonic point.  In combination with
sonic-point boundary conditions {\changed for stellar structure
  models, as they will be presented in a future work by
  Grassitelli et al.\ (submitted)}, such models could help to
better understand the interaction between stellar wind and envelope in
critical evolutionary phases near the Eddington limit, such as the LBV
phase or in phases directly preceding SN explosions.

\begin{acknowledgements}
  {\changed We thank the anonymous referee for his/her helpful comments,
    and W.-R.\ Hamann for providing the WRPLOT plotting
    software. G.G.\ thanks the Deutsche Forschunsgemeinschaft (DFG)
    for financial support under grant No.\ GR\,1717/5-1.}
\end{acknowledgements}

\begin{table*}
  \caption{ \label{tab:WRGAL} Empirical stellar and wind parameters for single
    H-free WN stars in the Galaxy.}
  {\small \begin{tabular}{lcccccccccccccccccc} \hline \hline \rule{0cm}{2.2ex}%
            ID  & $\log(L)$ & $T_\star$ & $R_\star$ & $M$ & $\log(\dot{M})$ & $\varv_\infty$ & $\varv_{\rm esc}$ & $\varv_\infty/\varv_{\rm esc}$ & $\eta$ & $\varv_\infty^{\rm\sc uv}$ \\
            \rule{0cm}{2.2ex}%
            WR  & $[L_\odot]$ & $[{\rm kK}]$ & $[R_\odot]$ & $[M_\odot]$ & $[\msunpyr]$ & $[\kms]$ & $[\kms]$ & & & $[\kms]$ \\
            \hline \rule{0cm}{2.2ex}%
            {}001  & 5.40 & 112.2 &  1.33 & 13.70 & -4.88 & 1900 & 1952 &  0.97 &  4.85 & 2223 \\
            {}002  & 5.45 & 141.3 &  0.89 & 14.57 & -5.47 & 1800 & 2461 &  0.73 &  1.06 & 2936 \\
            {}006  & 5.60 &  89.1 &  2.66 & 17.64 & -4.48 & 1700 & 1567 &  1.09 &  6.92 & 2251 \\
            {}046  & 5.80 & 112.2 &  2.11 & 23.08 & -5.25 & 2300 & 2013 &  1.14 &  1.00 & 2637 \\
            {}067  & 5.40 &  56.2 &  5.30 & 13.70 & -4.84 & 1500 &  971 &  1.54 &  4.28 &      \\
            {}074  & 5.40 &  56.2 &  5.30 & 13.70 & -4.75 & 1300 &  971 &  1.34 &  4.54 &      \\
            {}075  & 5.70 &  63.1 &  5.94 & 20.13 & -4.28 & 2300 & 1118 &  2.06 & 11.94 & 2280 \\
            {}115  & 5.65 &  50.1 &  8.90 & 18.83 & -4.72 & 1280 &  877 &  1.46 &  2.70 &      \\
            {}120  & 5.60 &  50.1 &  8.40 & 17.64 & -4.62 & 1225 &  874 &  1.40 &  3.59 &      \\
            {}134  & 5.60 &  63.1 &  5.29 & 17.64 & -4.63 & 1700 & 1106 &  1.54 &  4.89 & 2107 \\
            \hline \rule{0cm}{2.2ex}%
            {}007  & 5.45 & 112.2 &  1.41 & 14.57 & -4.92 & 1600 & 1953 &  0.82 &  3.34 & 2474 \\
            {}018  & 5.50 & 112.2 &  1.49 & 15.51 & -4.83 & 1800 & 1960 &  0.92 &  4.11 & 2472 \\
            {}020  & 5.60 &  63.1 &  5.29 & 17.64 & -4.96 & 1600 & 1104 &  1.45 &  2.17 &      \\
            {}034  & 5.50 &  63.1 &  4.72 & 15.51 & -4.94 & 1400 & 1096 &  1.28 &  2.49 &      \\
            {}036  & 5.30 &  89.1 &  1.88 & 12.14 & -4.51 & 1900 & 1546 &  1.23 & 14.50 &      \\
            {}037  & 5.50 & 100.0 &  1.88 & 15.51 & -4.75 & 2150 & 1749 &  1.23 &  5.87 &      \\
            {}044  & 5.55 &  79.4 &  3.16 & 16.53 & -5.20 & 1400 & 1385 &  1.01 &  1.21 & 2018 \\
            {}051  & 5.50 &  70.8 &  3.75 & 15.51 & -5.21 & 1500 & 1231 &  1.22 &  1.43 &      \\
            {}054  & 5.60 &  63.1 &  5.29 & 17.64 & -4.99 & 1500 & 1103 &  1.36 &  1.91 &      \\
            {}055  & 5.80 &  56.2 &  8.40 & 23.08 & -4.63 & 1200 & 1001 &  1.20 &  2.18 & 1675 \\
            {}061  & 5.40 &  63.1 &  4.21 & 13.70 & -4.87 & 1400 & 1092 &  1.28 &  3.72 & 1825 \\
            {}062  & 5.45 &  70.8 &  3.54 & 14.57 & -4.42 & 1800 & 1233 &  1.46 & 11.93 &      \\
            {}063  & 5.65 &  44.7 & 11.17 & 18.83 & -4.75 & 1700 &  781 &  2.18 &  3.36 &      \\
            {}084  & 5.65 &  50.1 &  8.90 & 18.83 & -4.78 & 1100 &  876 &  1.26 &  1.99 &      \\
            {}091  & 5.75 &  70.8 &  5.00 & 21.54 & -4.22 & 1700 & 1262 &  1.35 &  8.95 &      \\
            {}100  & 5.75 &  79.4 &  3.97 & 21.54 & -4.25 & 1600 & 1415 &  1.13 &  7.94 &      \\
            {}110  & 5.35 &  70.8 &  3.15 & 12.89 & -4.54 & 2300 & 1229 &  1.87 & 14.60 & 2410 \\
            {}123  & 6.05 &  44.7 & 17.71 & 33.24 & -4.09 &  970 &  826 &  1.17 &  3.46 & 1445 \\
            {}129  & 5.30 &  63.1 &  3.75 & 12.14 & -5.27 & 1320 & 1086 &  1.22 &  1.76 &      \\
            {}149  & 5.35 &  63.1 &  3.97 & 12.89 & -4.94 & 1300 & 1089 &  1.19 &  3.30 &      \\
            \hline \end{tabular}}
        \tablefoot{{\remark The indicated values are adopted from \citet{ham1:06},
            with terminal wind velocities $\varv_\infty^{\rm\sc uv}$ from \citet{nie1:02,nie1:04}. The luminosities of the stars in the upper panel are derived using known distances from cluster membership. The luminosities of the stars in the lower panel are estimated based on the $M_{\rm v}$ vs.\ spectral type calibration from \citet{ham1:06}, and thus more uncertain.}
          The mass-loss rates have been scaled down with $\dot{M}\propto 1/\sqrt{D}$ to match an 
          adopted clumping factor of $D=10$. In the original work $D=4$ was adopted.
          Masses are derived from the observed luminosities using the $M(L)$ relations for He stars from \citet{gra1:11}.}
\end{table*} 

\begin{table*}
  \caption{ \label{tab:WRLMC} {\changed Empirical} stellar and wind parameters for
    single H-free WN stars in the LMC.}
  {\small \begin{tabular}{lcccccccccccccccccc} \hline \hline \rule{0cm}{2.2ex}%
            ID  & $\log(L)$ & $T_\star$ & $R_\star$ & $M$ & $\log(\dot{M})$ & $\varv_\infty$ & $\varv_{\rm esc}$ & $\varv_\infty/\varv_{\rm esc}$ & $\eta$ & $\varv_\infty^{\rm\sc uv}$ \\
            \rule{0cm}{2.2ex}%
            BAT99  & $[L_\odot]$ & $[{\rm kK}]$ & $[R_\odot]$ & $[M_\odot]$ & $[\msunpyr]$ & $[\kms]$ & $[\kms]$ & & & $[\kms]$ \\
            \hline \rule{0cm}{2.2ex}%
            {}001 & 5.30 &  89.0 &  1.88 & 12.14 & -5.18 & 1600 & 1539 &  1.04 &  2.59 & 1470 \\
            {}002 & 5.37 & 141.0 &  0.81 & 13.21 & -5.28 & 1600 & 2446 &  0.65 &  1.76 & 2289 \\
            {}003 & 5.51 &  79.0 &  3.05 & 15.71 & -4.87 & 1600 & 1377 &  1.16 &  3.28 & 1607 \\
            {}005 & 5.45 & 141.0 &  0.89 & 14.57 & -5.22 & 1600 & 2454 &  0.65 &  1.69 &      \\
            {}007 & 5.84 & 158.0 &  1.11 & 24.41 & -4.47 & 1600 & 2844 &  0.56 &  3.81 & 2400 \\
            {}015 & 5.57 &  89.0 &  2.57 & 16.96 & -4.83 & 1600 & 1559 &  1.03 &  3.13 & 2082 \\
            {}017 & 5.69 &  67.0 &  5.21 & 19.86 & -4.97 & 1600 & 1182 &  1.35 &  1.72 & 2400 \\
            {}023 & 5.55 &  71.0 &  3.95 & 16.53 & -5.30 & 1600 & 1237 &  1.29 &  1.11 &      \\
            {}024 & 5.54 & 100.0 &  1.97 & 16.32 & -4.53 & 2400 & 1755 &  1.37 & 10.09 & 2294 \\
            {}026 & 5.62 &  71.0 &  4.28 & 18.10 & -4.80 & 1600 & 1246 &  1.28 &  3.01 & 2083 \\
            {}037 & 5.65 &  79.0 &  3.58 & 18.83 & -5.06 & 1600 & 1390 &  1.15 &  1.52 &      \\
            {}041 & 5.60 & 100.0 &  2.11 & 17.64 & -4.90 & 1300 & 1753 &  0.74 &  2.03 & 2241 \\
            {}046 & 5.44 &  63.0 &  4.42 & 14.39 & -5.23 & 1600 & 1090 &  1.47 &  1.69 & 2400 \\
            {}048 & 5.40 &  89.0 &  2.11 & 13.70 & -4.81 & 1600 & 1545 &  1.04 &  4.88 & 2242 \\
            {}051 & 5.30 &  89.0 &  1.88 & 12.14 & -5.18 & 1600 & 1539 &  1.04 &  2.59 & 2198 \\
            {}056 & 5.56 &  71.0 &  3.99 & 16.75 & -4.92 & 1600 & 1240 &  1.29 &  2.62 & 1470 \\
            {}057 & 5.40 &  79.0 &  2.68 & 13.70 & -4.95 & 1600 & 1370 &  1.17 &  3.50 & 2083 \\
            {}065 & 5.75 &  67.0 &  5.58 & 21.54 & -4.92 & 1600 & 1189 &  1.35 &  1.66 &      \\
            {}075 & 5.56 &  71.0 &  3.99 & 16.75 & -4.99 & 1600 & 1240 &  1.29 &  2.20 & 1841 \\
            {}086 & 5.33 &  71.0 &  3.06 & 12.58 & -5.46 & 1600 & 1225 &  1.31 &  1.26 &      \\
            {}088 & 5.80 & 112.0 &  2.12 & 23.08 & -4.81 & 1600 & 2006 &  0.80 &  1.95 &      \\
            {}094 & 5.80 & 141.0 &  1.33 & 23.08 & -4.51 & 1600 & 2526 &  0.63 &  3.88 & 1743 \\
            {}124 & 5.45 &  63.0 &  4.47 & 14.57 & -5.37 & 1600 & 1090 &  1.47 &  1.19 &      \\
            {}128 & 5.44 & 112.0 &  1.40 & 14.39 & -4.93 & 1600 & 1948 &  0.82 &  3.38 &      \\
            {}131 & 5.67 &  71.0 &  4.53 & 19.34 & -4.83 & 1600 & 1251 &  1.28 &  2.46 & 2289 \\
            {}132 & 5.58 &  79.0 &  3.30 & 17.18 & -4.67 & 1600 & 1385 &  1.16 &  4.46 & 1763 \\
            {}134 & 5.51 &  79.0 &  3.05 & 15.71 & -4.87 & 1600 & 1377 &  1.16 &  3.28 & 2081 \\
            \hline \end{tabular}}
        \tablefoot{{\remark The indicated values are adopted from \citet{hai1:14},
            with terminal wind velocities $\varv_\infty^{\rm\sc uv}$ from \citet{nie1:02,nie1:04}.}
          The mass-loss rates have been determined for a wind clumping factor of $D=10$.
          {\changed Masses are derived from the observed luminosities using the $M(L)$ relations for He stars from \citet{gra1:11}.}}
\end{table*} 

\bibliographystyle{aa}                                                         
\bibliography{astro}                                                           

\begin{thebibliography}{56}
\expandafter\ifx\csname natexlab\endcsname\relax\def\natexlab#1{#1}\fi

\bibitem[{{Abbott}(1980)}]{abb1:80}
{Abbott}, D.~C. 1980, \apj, 242, 1183

\bibitem[{{Castor} {et~al.}(1975){Castor}, {Abbott}, \& {Klein}}]{cas1:75}
{Castor}, J.~I., {Abbott}, D.~C., \& {Klein}, R.~I. 1975, \apj, 195, 157

\bibitem[{{Crowther} {et~al.}(2002){Crowther}, {Dessart}, {Hillier}, {Abbott},
  \& {Fullerton}}]{cro1:02}
{Crowther}, P.~A., {Dessart}, L., {Hillier}, D.~J., {Abbott}, J.~B., \&
  {Fullerton}, A.~W. 2002, \aap, 392, 653

\bibitem[{{Friend} \& {Castor}(1983)}]{fri1:83}
{Friend}, D.~B. \& {Castor}, J.~I. 1983, \apj, 272, 259

\bibitem[{{Gayley} {et~al.}(1995){Gayley}, {Owocki}, \& {Cranmer}}]{gay2:95}
{Gayley}, K.~G., {Owocki}, S.~P., \& {Cranmer}, S.~R. 1995, \apj, 442, 296

\bibitem[{{Gr{\"a}fener} \& {Hamann}(2005)}]{gra1:05}
{Gr{\"a}fener}, G. \& {Hamann}, W.-R. 2005, \aap, 432, 633

\bibitem[{{Gr{\"a}fener} \& {Hamann}(2008)}]{gra1:08}
{Gr{\"a}fener}, G. \& {Hamann}, W.-R. 2008, \aap, 482, 945

\bibitem[{{Gr\"afener} {et~al.}(1998){Gr\"afener}, {Hamann}, {Hillier}, \&
  {Koesterke}}]{gra1:98}
{Gr\"afener}, G., {Hamann}, W.-R., {Hillier}, D.~J., \& {Koesterke}, L. 1998,
  \aap, 329, 190

\bibitem[{{Gr{\"a}fener} {et~al.}(2008){Gr{\"a}fener}, {Hamann}, \&
  {Todt}}]{gra5:08}
{Gr{\"a}fener}, G., {Hamann}, W.-R., \& {Todt}, H. 2008, in Astronomical
  Society of the Pacific Conference Series, Vol. 391, Hydrogen-Deficient Stars,
  ed. A.~{Werner} \& T.~{Rauch}, 99

\bibitem[{Gr\"afener {et~al.}(2002)Gr\"afener, Koesterke, \& Hamann}]{gra1:02}
Gr\"afener, G., Koesterke, L., \& Hamann, W.-R. 2002, \aap, 387, 244

\bibitem[{{Gr{\"a}fener} {et~al.}(2012){Gr{\"a}fener}, {Owocki}, \&
  {Vink}}]{gra1:12}
{Gr{\"a}fener}, G., {Owocki}, S.~P., \& {Vink}, J.~S. 2012, \aap, 538, A40

\bibitem[{{Gr{\"a}fener} \& {Vink}(2013)}]{gra2:13}
{Gr{\"a}fener}, G. \& {Vink}, J.~S. 2013, \aap, 560, A6

\bibitem[{{Gr{\"a}fener} {et~al.}(2011){Gr{\"a}fener}, {Vink}, {de Koter}, \&
  {Langer}}]{gra1:11}
{Gr{\"a}fener}, G., {Vink}, J.~S., {de Koter}, A., \& {Langer}, N. 2011, \aap,
  535, A56

\bibitem[{{Grassitelli} {et~al.}(2016){Grassitelli}, {Chen{\'e}}, {Sanyal},
  {Langer}, {St-Louis}, {Bestenlehner}, \& {Fossati}}]{gra2:16}
{Grassitelli}, L., {Chen{\'e}}, A.-N., {Sanyal}, D., {et~al.} 2016, \aap, 590,
  A12

\bibitem[{{Grevesse} \& {Noels}(1993)}]{gre1:93}
{Grevesse}, N. \& {Noels}, A. 1993, in Origin and Evolution of the Elements,
  ed. N.~{Prantzos}, E.~{Vangioni-Flam}, \& M.~{Casse}, 15--25

\bibitem[{{Hainich} {et~al.}(2014){Hainich}, {R{\"u}hling}, {Todt}, {Oskinova},
  {Liermann}, {Gr{\"a}fener}, {Foellmi}, {Schnurr}, \& {Hamann}}]{hai1:14}
{Hainich}, R., {R{\"u}hling}, U., {Todt}, H., {et~al.} 2014, \aap, 565, A27

\bibitem[{{Hamann} \& {Gr{\"a}fener}(2003)}]{ham1:03}
{Hamann}, W.-R. \& {Gr{\"a}fener}, G. 2003, \aap, 410, 993

\bibitem[{{Hamann} \& {Gr{\"a}fener}(2004)}]{ham1:04}
{Hamann}, W.-R. \& {Gr{\"a}fener}, G. 2004, \aap, 427, 697

\bibitem[{{Hamann} {et~al.}(2006){Hamann}, {Gr{\"a}fener}, \&
  {Liermann}}]{ham1:06}
{Hamann}, W.-R., {Gr{\"a}fener}, G., \& {Liermann}, A. 2006, \aap, 457, 1015

\bibitem[{{Heger} \& {Langer}(1996)}]{heg1:96}
{Heger}, A. \& {Langer}, N. 1996, \aap, 315, 421

\bibitem[{{Iglesias} \& {Rogers}(1996)}]{igl1:96}
{Iglesias}, C.~A. \& {Rogers}, F.~J. 1996, \apj, 464, 943

\bibitem[{{Ishii} {et~al.}(1999){Ishii}, {Ueno}, \& {Kato}}]{ish1:99}
{Ishii}, M., {Ueno}, M., \& {Kato}, M. 1999, \pasj, 51, 417

\bibitem[{{Jiang} {et~al.}(2015){Jiang}, {Cantiello}, {Bildsten}, {Quataert},
  \& {Blaes}}]{jia1:15}
{Jiang}, Y.-F., {Cantiello}, M., {Bildsten}, L., {Quataert}, E., \& {Blaes}, O.
  2015, \apj, 813, 74

\bibitem[{{Joss} {et~al.}(1973){Joss}, {Salpeter}, \& {Ostriker}}]{jos1:73}
{Joss}, P.~C., {Salpeter}, E.~E., \& {Ostriker}, J.~P. 1973, \apj, 181, 429

\bibitem[{{Lamers} \& {Cassinelli}(1999)}]{lam1:99}
{Lamers}, H.~J.~G.~L.~M. \& {Cassinelli}, J.~P. 1999, {Introduction to Stellar
  Winds} (Cambridge University Press)

\bibitem[{{Lamers} \& {Nugis}(2002)}]{lam1:02}
{Lamers}, H.~J.~G.~L.~M. \& {Nugis}, T. 2002, \aap, 395, L1

\bibitem[{{Lamers} {et~al.}(1995){Lamers}, {Snow}, \& {Lindholm}}]{lam1:95}
{Lamers}, H.~J.~G.~L.~M., {Snow}, T.~P., \& {Lindholm}, D.~M. 1995, \apj, 455,
  269

\bibitem[{{Langer}(1989)}]{lan1:89}
{Langer}, N. 1989, \aap, 210, 93

\bibitem[{{Lucy}(1971)}]{luc1:71}
{Lucy}, L.~B. 1971, \apj, 163, 95

\bibitem[{{Lucy}(1976)}]{luc1:76}
{Lucy}, L.~B. 1976, \apj, 205, 482

\bibitem[{{Lucy} \& {Abbott}(1993)}]{luc1:93}
{Lucy}, L.~B. \& {Abbott}, D.~C. 1993, \apj, 405, 738

\bibitem[{{Netzer} \& {Elitzur}(1993)}]{net1:93}
{Netzer}, N. \& {Elitzur}, M. 1993, \apj, 410, 701

\bibitem[{{Niedzielski} {et~al.}(2004){Niedzielski}, {Nugis}, \&
  {Skorzynski}}]{nie1:04}
{Niedzielski}, A., {Nugis}, T., \& {Skorzynski}, W. 2004, \actaa, 54, 405

\bibitem[{{Niedzielski} \& {Skorzynski}(2002)}]{nie1:02}
{Niedzielski}, A. \& {Skorzynski}, W. 2002, Acta Inf., 52, 81

\bibitem[{{Nugis} \& {Lamers}(2000)}]{nug1:00}
{Nugis}, T. \& {Lamers}, H.~J.~G.~L.~M. 2000, \aap, 360, 227

\bibitem[{{Nugis} \& {Lamers}(2002)}]{nug1:02}
{Nugis}, T. \& {Lamers}, H.~J.~G.~L.~M. 2002, \aap, 389, 162

\bibitem[{{Oskinova} {et~al.}(2007){Oskinova}, {Hamann}, \&
  {Feldmeier}}]{osk1:07}
{Oskinova}, L.~M., {Hamann}, W.-R., \& {Feldmeier}, A. 2007, \aap, 476, 1331

\bibitem[{{Owocki}(2008)}]{owo1:08}
{Owocki}, S.~P. 2008, in Clumping in Hot-Star Winds, ed. W.-R. {Hamann},
  A.~{Feldmeier}, \& L.~M. {Oskinova}, 121

\bibitem[{{Owocki} \& {Gayley}(1997)}]{owo1:97}
{Owocki}, S.~P. \& {Gayley}, K.~G. 1997, in Astronomical Society of the Pacific
  Conference Series, Vol. 120, Luminous Blue Variables: Massive Stars in
  Transition, ed. {A.~Nota \& H.~Lamers}, 121

\bibitem[{{Owocki} {et~al.}(2004){Owocki}, {Gayley}, \& {Shaviv}}]{owo1:04}
{Owocki}, S.~P., {Gayley}, K.~G., \& {Shaviv}, N.~J. 2004, \apj, 616, 525

\bibitem[{{Petrov} {et~al.}(2016){Petrov}, {Vink}, \& {Gr{\"a}fener}}]{pet1:16}
{Petrov}, B., {Vink}, J.~S., \& {Gr{\"a}fener}, G. 2016, \mnras, 458, 1999

\bibitem[{{Petrovic} {et~al.}(2006){Petrovic}, {Pols}, \& {Langer}}]{pet1:06}
{Petrovic}, J., {Pols}, O., \& {Langer}, N. 2006, \aap, 450, 219

\bibitem[{{Ro} \& {Matzner}(2016)}]{ro1:16}
{Ro}, S. \& {Matzner}, C.~D. 2016, \apj, 821, 109

\bibitem[{{Sander} {et~al.}(2012){Sander}, {Hamann}, \& {Todt}}]{san1:12}
{Sander}, A., {Hamann}, W.-R., \& {Todt}, H. 2012, \aap, 540, A144

\bibitem[{{Sanyal} {et~al.}(2015){Sanyal}, {Grassitelli}, {Langer}, \&
  {Bestenlehner}}]{san1:15}
{Sanyal}, D., {Grassitelli}, L., {Langer}, N., \& {Bestenlehner}, J.~M. 2015,
  \aap, 580, A20

\bibitem[{{Shaviv}(1998)}]{sha1:98}
{Shaviv}, N.~J. 1998, \apjl, 494, L193

\bibitem[{{Smith} {et~al.}(2017){Smith}, {Gotberg}, \& {de Mink}}]{smi1:17}
{Smith}, N., {Gotberg}, Y., \& {de Mink}, S.~E. 2017, ArXiv e-prints
  [\eprint[arXiv]{1704.03516}]

\bibitem[{{Sundqvist} {et~al.}(2014){Sundqvist}, {Puls}, \& {Owocki}}]{sun1:14}
{Sundqvist}, J.~O., {Puls}, J., \& {Owocki}, S.~P. 2014, \aap, 568, A59

\bibitem[{{Vink} \& {de Koter}(2002)}]{vin1:02}
{Vink}, J.~S. \& {de Koter}, A. 2002, \aap, 393, 543

\bibitem[{{Vink} \& {de Koter}(2005)}]{vin1:05}
{Vink}, J.~S. \& {de Koter}, A. 2005, \aap, 442, 587

\bibitem[{{Vink} {et~al.}(1999){Vink}, {de Koter}, \& {Lamers}}]{vin1:99}
{Vink}, J.~S., {de Koter}, A., \& {Lamers}, H.~J.~G.~L.~M. 1999, \aap, 350, 181

\bibitem[{{Vink} {et~al.}(2000){Vink}, {de Koter}, \& {Lamers}}]{vin1:00}
{Vink}, J.~S., {de Koter}, A., \& {Lamers}, H.~J.~G.~L.~M. 2000, \aap, 362, 295

\bibitem[{{Vink} \& {Gr{\"a}fener}(2012)}]{vin1:12}
{Vink}, J.~S. \& {Gr{\"a}fener}, G. 2012, \apjl, 751, L34

\bibitem[{{Vink} {et~al.}(2011){Vink}, {Muijres}, {Anthonisse}, {de Koter},
  {Gr{\"a}fener}, \& {Langer}}]{vin1:11}
{Vink}, J.~S., {Muijres}, L.~E., {Anthonisse}, B., {et~al.} 2011, \aap, 531,
  A132

\bibitem[{{Yoon}(2017)}]{yoo1:17}
{Yoon}, S.-C. 2017, \mnras, 470, 3970

\bibitem[{{Zapartas} {et~al.}(2017){Zapartas}, {de Mink}, {Van Dyk}, {Fox},
  {Smith}, {Bostroem}, {de Koter}, {Filippenko}, {Izzard}, {Kelly}, {Neijssel},
  {Renzo}, \& {Ryder}}]{zap1:17}
{Zapartas}, E., {de Mink}, S.~E., {Van Dyk}, S.~D., {et~al.} 2017, \apj, 842,
  125

\end{thebibliography}

\end{document}